\documentclass[twocolumn]{aastex631}
\usepackage{amsmath}

\newcommand\kms{km$\,$s$^{-1}$}
\newcommand\Msol{M$_{\odot}$}

\newcommand{\hi}{H\,{\sc i}}
\newcommand{\hii}{H\,{\sc ii}}

\submitjournal{AAS Journals}

\shorttitle{GC systems of field UDGs}
\shortauthors{Jones et al.}
\graphicspath{{./}{figures/}}

\begin{document}

\title{Gas-rich, field ultra-diffuse galaxies host few globular clusters}

\correspondingauthor{Michael G. Jones}
\email{jonesmg@arizona.edu}

\author[0000-0002-5434-4904]{Michael G. Jones}
\affiliation{Steward Observatory, University of Arizona, 933 North Cherry Avenue, Rm. N204, Tucson, AZ 85721-0065, USA}

\author[0000-0001-8855-3635]{Ananthan Karunakaran}
\affiliation{Instituto de Astrof\'{i}sica de Andaluc\'{i}a (CSIC), Glorieta de la Astronom\'{i}a, 18008 Granada, Spain}

\author[0000-0001-8354-7279]{Paul Bennet}
\affiliation{Space Telescope Science Institute, 3700 San Martin Drive, Baltimore, MD 21218, USA}

\author[0000-0003-4102-380X]{David J. Sand}
\affiliation{Steward Observatory, University of Arizona, 933 North Cherry Avenue, Rm. N204, Tucson, AZ 85721-0065, USA}

\author[0000-0002-0956-7949]{Kristine Spekkens}
\affiliation{Department of Physics and Space Science, Royal Military College of Canada P.O. Box 17000, Station Forces Kingston, ON K7K 7B4, Canada}
\affiliation{Department of Physics, Engineering Physics and Astronomy, Queen’s University, Kingston, ON K7L 3N6, Canada}

\author[0000-0001-9649-4815]{Bur\c{c}in Mutlu-Pakdil}
\affil{Department of Physics and Astronomy, Dartmouth College, Hanover, NH 03755, USA}

\author[0000-0002-1763-4128]{Denija Crnojevi\'{c}}
\affil{University of Tampa, 401 West Kennedy Boulevard, Tampa, FL 33606, USA}

\author[0000-0001-9165-8905]{Steven Janowiecki}
\affil{University of Texas, Hobby-Eberly Telescope, McDonald Observatory, TX 79734, USA}

\author[0000-0001-8849-7987]{Lukas Leisman}
\affil{Department of Astronomy, University of Illinois, 1002 W. Green St., Urbana, IL 61801, USA}
\affil{Department of Physics and Astronomy, Valparaiso University, 1610 Campus Drive East, Valparaiso, IN 46383, USA}

\author[0000-0001-8245-779X]{Catherine E. Fielder}
\affiliation{Steward Observatory, University of Arizona, 933 North Cherry Avenue, Rm. N204, Tucson, AZ 85721-0065, USA}



\begin{abstract}

We present Hubble Space Telescope imaging of 14 gas-rich, low surface brightness galaxies in the field at distances of 25-36~Mpc, with mean effective radii and $g$-band central surface brightnesses of 1.9~kpc and 24.2 mag arcsec$^{-2}$. Nine meet the standard criteria to be considered ultra-diffuse galaxies (UDGs). An inspection of point-like sources brighter than the turnover magnitude of the globular cluster luminosity function and within twice the half-light radii of each galaxy reveals that, unlike those in denser environments, gas-rich, field UDGs host very few old globular clusters (GCs).  Most of the targets (nine) have zero candidate GCs, with the remainder having one or two candidates each. These findings are broadly consistent with expectations for normal dwarf galaxies of similar stellar mass.
This rules out gas-rich, field UDGs as potential progenitors of the GC-rich UDGs that are typically found in galaxy clusters. However, some in galaxy groups may be directly accreted from the field.
In line with other recent results, this strongly suggests that there must be at least two distinct formation pathways for UDGs, and that this sub-population is simply an extreme low surface brightness extension of the underlying dwarf galaxy population. The root cause of their diffuse stellar distributions remains unclear, but the formation mechanism appears to only impact the distribution of stars (and potentially dark matter), without strongly impacting the distribution of neutral gas, the overall stellar mass, or the number of GCs.

\end{abstract}



\section{Introduction} 
\label{sec:intro}

The study of low surface brightness (LSB) galaxies, including dwarf galaxies, has a long history \citep{Sandage+1984,Impey+1988,Thompson+1993,Jerjen+2000,Conselice+2003,Mieske+2007}. However, the abundance of the most extreme LSB galaxies was not fully appreciated until relatively recently when hundreds of ultra-diffuse galaxies (UDGs) were identified in nearby galaxy clusters \citep{vanDokkum+2015,Mihos+2015,Koda+2015,Yagi+2016,vanderBurg+2016,Wittmann+2017,Venhola+2017,Zaritsky+2019}. Soon large samples of UDGs were identified in all environments from those in clusters, to galaxy groups \citep{vanderBurg+2017,Trujillo+2017,Roman+2017,Spekkens+2018,Bennet+2018}, and to the field \citep{Leisman+2017,Janowiecki+2019,Roman+2019,Prole+2019,Karunakaran+2020b}.
These findings led to a number of potential formation models including both internal mechanisms \citep[e.g.][]{Amorisco+2016,DiCintio+2017,Chan+2018} and those relying on external, environmental effects \citep{Conselice+2018,Carleton+2019,Tremmel+2020}. It has also been hypothesized that UDGs may be the result of a combination of internal and external mechanisms, either acting jointly \citep{Martin+2019,Jackson+2021}, or with different pathways being responsible for different subsets of the UDG population \citep{Papastergis+2017,Pandya+2018,Jiang+2019,Liao+2019,Wright+2021,Buzzo+2022}. These mechanisms range from early truncated growth, star formation feedback, intrinsic halo properties, tidal rarification, and mergers \citep[for a recent summary of these proposed mechanisms, see][]{Jones+2021}. While a number of works have argued that UDGs may have multiple formation pathways, it is still unclear whether UDGs in the field are directly related to those in denser environments. Could field UDGs be the progenitors of UDGs in denser environments, or are these largely distinct populations that formed via unrelated processes?

A key metric related to the early stages of formation of a galaxy is the number of old globular clusters (GCs) that it hosts. The richness of a galaxy's GC system is strongly correlated with its total mass \citep[e.g.][]{Blakeslee+1997,Harris+2013,Zaritsky+2022}, for which dark matter (DM) halo mass, stellar masses, and luminosity (in order of decreasing linearity and tightness of the relation) may all be used as proxies. GCs therefore offer a means to probe the DM halo masses of UDGs \citep[which appear to follow the established relation,][]{Harris+2017}, but also a means to compare subsets of the UDG population that are otherwise similar in terms of their surface brightness and stellar mass. 

Some investigations of the GC systems of UDGs in clusters have found them to be extraordinarily rich \citep{Beasley+2016a,Peng+2016,vanDokkum+2016,Beasley+2016b,vanDokkum+2017}, while others have argued their GC systems are less extreme \citep{Amorisco+2018,Somalwar+2020,Forbes+2020,Lim+2020,Saifollahi+2021}. However, even with the lower GC count estimates, UDGs still host richer GC systems on average than other dwarf galaxies of equivalent luminosity or stellar mass, but they correspond to dwarf-mass DM halos (e.g. $10^{10}-10^{11.5}$ \Msol), not Milky Way-mass halos (e.g. $\sim$10$^{12}$ \Msol).

\citet{Jones+2021} used Hubble Space Telescope (HST) observations to identify GC candidates (GCCs) in two group UDGs that appeared to have (tidal) stellar streams connecting them to their respective hosts \citep{Bennet+2018}. Unlike most UDGs in clusters, these UDGs appeared to host a small number of GCs, roughly in line with expectations for typical dwarf galaxies. This strongly suggested that these were previously regular dwarfs that were ``puffed up'' by tidal interactions with their hosts, after falling into a group \citep[e.g.][]{Carleton+2019}. This would make them distinct from cluster UDGs, but also UDGs that became ultra-diffuse while in the field, presumably via some internal mechanism. However, a significant caveat to this finding still remains: we are still largely ignorant of the properties of the GC systems of field UDGs. They may also be consistent with those of typical dwarf galaxies, in which case it would be less clear whether group UDGs such as those identified by \citet{Bennet+2018} are truly distinct from those in the field. They plausibly could have already been ultra-diffuse prior to falling into their current groups, and the evidence of tidal interactions may have no bearing on their status as UDGs. Equally, it may be possible that field UDGs are instead GC-rich and represent the progenitors of UDGs in cluster environments.

In this work we address this missing information by performing a census of the GCCs in 14 UDGs and LSB galaxies in the field, using HST Wide Field Camera 3 (WFC3) snapshot observations. The paper is organized as follows. In \S\ref{sec:obs} we describe our target sample and observational strategy. In \S\ref{sec:results} we explain our approach to selecting GCCs and present the resulting GC counts. In \S\ref{sec:discuss} we discuss the interpretation of these results and present our conclusions in \S\ref{sec:conclude}.

\section{Sample \& Observations}
\label{sec:obs}

\begin{figure}
    \centering
    \includegraphics[width=\columnwidth]{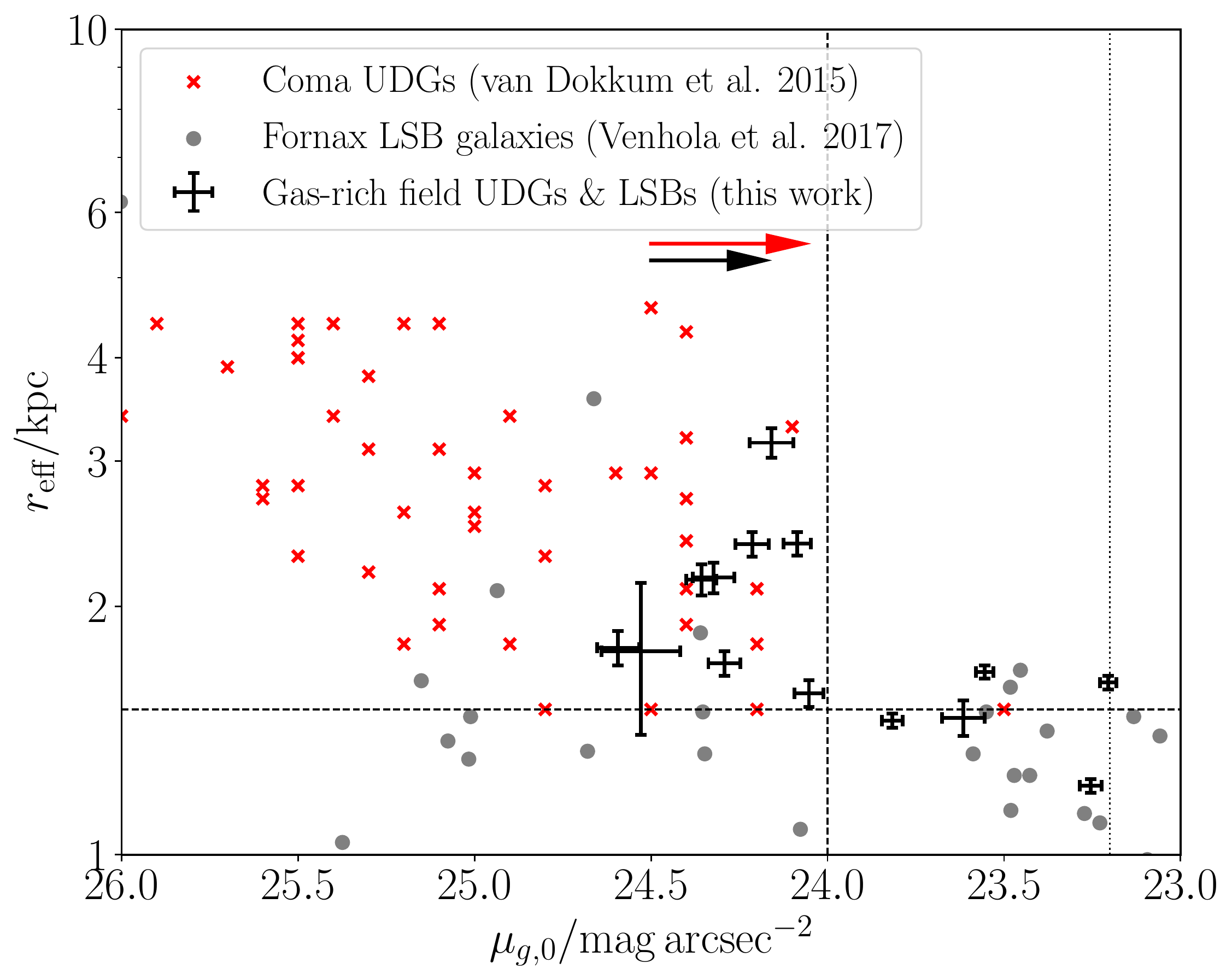}
    \caption{Effective radius versus central surface brightness for our target sample (black error bars) compared to the \citet{vanDokkum+2015} UDGs in Coma (red crosses) and LSB galaxies in Fornax \citep[grey circles,][]{Venhola+2017}. The horizontal dashed line indicates the standard $r_\mathrm{eff} = 1.5$~kpc cutoff for UDGs. The vertical dashed line at $\mu_\mathrm{g,0} = 24$ corresponds to the UDG surface brightness criterion of \citet{vanDokkum+2015}, while the more relaxed limit of \citet{vanderBurg+2016} is approximated by the dotted line at $\mu_\mathrm{g,0} = 23.2$ (for a typical color of $g-r = 0.3$). Nine of our targets meet the more stringent criteria, while 11 meet the more relaxed criteria. The horizontal red and black arrows indicate how the surface brightness values of the \citet{vanDokkum+2015} UDGs and our sample would shift (on average, based on their mean $g-r$ colors) if the $r$-band central surface brightness was used instead of $g$-band. As our objects are bluer their surface brightness does not increase as much between $g$ and $r$-band.
    }
    \label{fig:sample}
\end{figure}

The UDGs targeted in this work are drawn from the \hi-bearing UDGs sample of \citet{Janowiecki+2019}. These are UDGs originally detected through their \hi \ line emission in the Arecibo Legacy Fast ALFA (Arecibo L-band Feed Array) survey \citep[ALFALFA,][]{Giovanelli+2005,Haynes+2011,Haynes+2018} and subsequently identified as UDGs based on their LSB counterparts in Sloan Digital Sky Survey \citep[SDSS,][]{York+2000} images. This catalog is an expansion and revision of the initial ALFALFA-based catalog of \citet{Leisman+2017}. Part of this revision was to drop the explicit requirement for objects to be isolated from other, larger galaxies \citep[as was the case in][]{Leisman+2017}, but most of the expansion of the sample is simply the result of including more of the full footprint of ALFALFA, which was unavailable when the original catalog was published. 

Despite no explicit requirement for isolation, \citet{Janowiecki+2019} demonstrate that \hi-bearing UDGs reside in environments typical of other similar mass, gas-rich galaxies in the ALFALFA survey, which are generally low-mass centrals in their own halos \citep{Guo+2017}. Furthermore, none of our targets were matched to known groups by \citet{Jones+2020}. Thus, these \hi-bearing UDGs are bona fide field objects, not satellites of larger galaxies or groups. This distinguishes this sample from other field UDG samples, such as those identified in the ``Mass Assembly of early Type gaLAxies with their fine Structures" (MATLAS) survey \citep{Habas+2020,Marleau+2021}, whose members, though in low density environments, are still mostly satellites.

The detection of \hi \ is a prerequisite for the identification of these objects, meaning that, unlike most UDG samples, which are based purely on photometry, all have known spectroscopic redshifts. Furthermore, the parent sample \citep{Janowiecki+2019} does not consider candidates within $\sim$25~Mpc, which removes those with the largest fractional uncertainties (due to peculiar velocities) on their redshift-based distance estimates. A maximum distance limit of 120~Mpc was also applied, as beyond this limit the projected distance corresponding to Arecibo's $\sim$3.5\arcmin \ beam can complicate the robust identification of the optical counterparts of \hi \ detections, especially for LSB objects.

The UDGs and LSB galaxies in our observed sample are shown in Figure \ref{fig:sample} relative to UDGs in a cluster environment.

\subsection{HST targets and observations}

In order to quantify the GC populations of field UDGs we proposed for a cycle 29 HST snapshot imaging program with WFC3. All the UDGs within 40~Mpc from \citet{Janowiecki+2019}, a total of 21 objects, were submitted for snapshot imaging (HST-16758, PI: M.~Jones). The limit of 40~Mpc was to ensure that even for the most distant targets the turnover in the GCLF could be reached at high signal-to-noise ratio ($\mathrm{SNR}>5$) in two filters in a single orbit.
A total of 15 of these targets were observed during cycle 29. Each target was observed for a total of 1000~s in two exposures in the F555W filter and 750~s in two exposures in F814W. Unfortunately, the observations of AGC~242019 lost tracking during the F555W exposures and the images were unusable, resulting in a sample of 14 targets (Table \ref{tab:targets}).
The false color images from the two filters combined are shown for each target in Figure \ref{fig:HST_images}.\footnote{These images are available in the Mikulski Archive for Space Telescopes, \dataset[10.17909/6n6q-ke17]{https://doi.org/10.17909/6n6q-ke17}.}

\begin{table*}
\centering
\caption{Target sample of \hi-bearing UDGs \& LSB galaxies}
\begin{tabular}{cccccccccccc}
\hline\hline
AGC &   RA &  Dec &  $cz$ &  Dist &  $\log \frac{M_\mathrm{HI}}{\mathrm{M_\odot}}$ &  $M_g$ & $g-r$ & $\mu_{0,g}$ &  $r_\mathrm{eff}$  & $\log \frac{M_\ast}{\mathrm{M_\odot}}$ & $N_\mathrm{GCC}$      \\
 & deg & deg & $\mathrm{km\,s^{-1}}$ & Mpc & & mag & mag & mag arcsec$^{-2}$ & kpc & & \\
\hline
100288 & 7.9387 & 2.7161 & 2376 & 32.9 & 8.50 & -15.1 & 0.3 & $24.16 \pm 0.04$ & $2.38 \pm 0.08$ & $7.81 \pm 0.12$ & 0 \\
102375 & 3.7491 & 2.5800 & 2567 & 35.8 & 8.23 & -14.3 & 0.4 & $23.88 \pm 0.03$ & $1.45 \pm 0.03$ & $7.63 \pm 0.10$ & 0 \\
103435 & 3.8379 & 1.0750 & 2048 & 28.5 & 8.28 & -14.3 & 0.2 & $24.36 \pm 0.05$ & $1.71 \pm 0.06$ & $7.18 \pm 0.17$ & 2 \\
114959 & 22.8283 & 11.7150 & 2555 & 34.6 & 8.51 & -14.5 & 0.2 & $23.82 \pm 0.06$ & $1.47 \pm 0.07$ & $7.42 \pm 0.14$ & 0 \\
115292 & 27.9149 & 17.2881 & 2564 & 34.5 & 8.32 & -14.8 & 0.3 & $23.70 \pm 0.03$ & $1.67 \pm 0.03$ & $7.73 \pm 0.11$ & 0 \\
124634 & 35.5229 & 18.4089 & 2290 & 30.6 & 8.48 & -14.6 & 0.4 & $23.79 \pm 0.02$ & $1.62 \pm 0.03$ & $7.78 \pm 0.09$ & 0 \\
181474 & 121.5521 & 15.5040 & 1986 & 30.4 & 8.45 & -14.5 & 0.2 & $24.44 \pm 0.04$ & $2.16 \pm 0.09$ & $7.37 \pm 0.15$ & 0 \\
189298 & 133.1097 & 30.7260 & 2113 & 31.8 & 8.19 & -14.1 & 0.3 & $24.14 \pm 0.04$ & $1.57 \pm 0.06$ & $7.47 \pm 0.11$ & 1 \\
191708 & 145.1132 & 0.0442 & 1887 & 29.3 & 8.34 & -15.0 & 0.4 & $24.36 \pm 0.05$ & $2.38 \pm 0.08$ & $7.97 \pm 0.09$ & 1 \\
198686 & 135.9513 & 31.7847 & 1980 & 30.0 & 8.25 & -14.6 & 0.4 & $23.31 \pm 0.03$ & $1.21 \pm 0.02$ & $7.69 \pm 0.11$ & 0 \\
201993 & 153.9986 & 6.8042 & 1620 & 25.8 & 8.41 & -14.8 & 0.3 & $24.38 \pm 0.06$ & $2.17 \pm 0.09$ & $7.62 \pm 0.13$ & 0 \\
258471 & 230.6593 & 5.8299 & 1796 & 28.9 & 7.94 & -12.8 & 0.3 & $24.67 \pm 0.11$ & $1.77 \pm 0.37$ & $6.82 \pm 0.13$ & 1 \\
312297 & 328.2307 & 13.5611 & 1721 & 27.2 & 8.20 & -13.7 & 0.4 & $24.83 \pm 0.06$ & $1.78 \pm 0.09$ & $7.40 \pm 0.10$ & 1 \\
749387 & 126.7348 & 24.7290 & 2289 & 34.4 & 8.11 & -15.5 & 0.4 & $24.34 \pm 0.06$ & $3.15 \pm 0.13$ & $8.22 \pm 0.08$ & 0 \\
\hline
\end{tabular}\\
Columns: 1) AGC number as in \citet{Haynes+2018}. 2) Right ascension in decimal degrees. 3) Declination in decimal degrees. 4) Heliocentric redshift of the \hi \ line \citep{Haynes+2018}. 5) Distance estimated from ALFALFA flow model \citep{Masters2005}. 6) Logarithm of \hi mass in solar masses \citep{Haynes+2018}. 7) Absolute $g$-band magnitude from DECaLS photometry. 8) $g-r$ color. 9) Central $g$-band surface brightness. 10) Effective radius in $g$-band. 11) Logarithm of stellar mass (Appendix \ref{sec:stellar_masses}). 12) Number of globular cluster candidates (uncorrected).
\label{tab:targets}
\end{table*}

\subsection{Revision of photometry with DECaLS}
\label{sec:DECaLS_phot}

The optical photometry of the parent \hi-bearing UDGs sample was measured from SDSS images \citep{Leisman+2017,Janowiecki+2019} and size and surface brightness criteria were set to approximately match those of \citet{vanderBurg+2016} for the full sample, and those of \citet{vanDokkum+2015} for a more restricted sample. Deeper DECaLS \citep[Dark Energy Camera Legacy Survey,][]{Dey+2019} images are now available for all our targets and we have therefore revised their photometry (Figure \ref{fig:sample}).

The photometry for these UDGs is performed using \texttt{AutoProf} \citep{Stone+2021}, a flexible, non-parametric fitting \texttt{Python} package that incorporates machine learning methods (i.e.\ regularization) to improve upon previous fitting implementations. We retrieve 5\arcmin \ $g$ and $r-$band cutouts for each of the \hi-bearing UDGs from DECaLS. Masks for these systems are generated using \texttt{DeepScan} \citep{Prole+2018} and used in the surface brightness profile extraction. We take advantage of \texttt{AutoProf}'s flexibility to measure the surface brightness profiles of our UDGs within circular apertures in a similar manner to \citet{Leisman+2017} and \citet{Janowiecki+2019}.\ We then fit the extracted surface brightness profiles with an exponential function to estimate their central surface brightness and effective radii (Table \ref{tab:targets}) with uncertainties on these quantities estimated via bootstrap re-sampling.

With these deeper data, the uncertainties on the photometry have been significantly reduced and most objects have moved to slightly lower (brighter) $\mu_{g,0}$ values (central $g$-band surface brightness). Nine of the 14 targets meet the \citet{vanDokkum+2015} criteria for UDGs, while 11 meet the \citet{vanderBurg+2016} criteria, and the remaining three are LSB galaxies near the border of these definitions (Figure \ref{fig:sample}). Although the \citet{vanDokkum+2015} definition is more widely used, the \citet{vanderBurg+2016} definition may be more appropriate for blue UDGs. Firstly, they are blue because of recent star formation events which will brighten their magnitude more in $g$-band than $r$-band. Thus, the $r$-band surface brightness is more representative of the total stellar content \citep[cf.][]{Li+2022}. Secondly, using central surface brightness makes most sense for galaxies with smooth light distributions, which can be accurately modeled with S\'{e}rsic profiles. Most gas-rich, field UDGs have decidedly irregular and clumpy morphologies that only loosely follow an exponential profile. However, on average they do fall into the same extreme LSB regime as redder, smoother UDGs. 

As the exact thresholds of surface brightness and size used to demarcate UDGs versus LSB galaxies is largely arbitrary, we will consider this small sample as a whole for the remainder of this work. However, we note that all of the qualitative findings would be unchanged if the sample were to be restricted to the nine objects that meet the most stringent criteria.

We also used the HST images to verify our new photometric measurements from DECaLS. We model each target galaxy using \texttt{GALFIT} \citep{Peng+2002}, following the procedure from \cite{Bennet+2017}. These photometric fits were consistent with those from DECaLS, but generally found smaller radii, fainter integrated magnitudes, and significantly higher uncertainties. 
Thus, we elect to rely on the DECaLS results for the remainder of this work.

\section{Globular cluster candidates}
\label{sec:results}

All individual WFC3 exposures (in both filters) were aligned and a combined source catalog was extracted for each target using \texttt{DOLPHOT} \citep{Dolphin2000,Dolphin2016}. To select GCCs we begin by restricting the catalog to point-like sources (object type 1 and 2) with no photometry flags in either filter. Next a SNR minimum of 5 is enforced. Sources with more than 0.5 mag of additional flux (in the two filters combined) due to crowding are removed. The absolute sharpness value is required to be less than 0.25 in order to remove highly extended sources. Finally, a very loose roundness threshold of $<3$ is enforced, which helps to remove any remaining diffraction spikes or other highly elongated features. The F555W and F814W magnitude of the remaining sources are corrected for Galactic extinction following \citet{Schlafly+2011}. The values of E(B-V) are all less than 0.2, and most are less than 0.05. The corrected magnitudes are then converted to V and I (Vega) magnitudes following \citet{Harris2018}.

This results in a catalog of high SNR, point-like objects that are concentrated around our target galaxy in each HST image. However, many of these sources correspond to young star forming regions, not old GCs. Confusion between young star clusters and GCs is a well known issue for the identification of GCs around late-type and irregular dwarfs. We adopt a simple color cut of $V-I > 0.85$ to remove young clusters.  \citet{Seth+2004} argue that any star cluster with a single stellar population that is older than 1~Gyr will have a color $V-I > 0.85$, regardless of metallicity. We independently verified this cut using the PAdova and TRieste Stellar Evolution Code \citep{Bressan+2012}. We also set an upper limit on the color of $V-I < 1.5$, as almost no GCs are redder than this \citep{Brodie+2006}.

To prevent contamination from bright stars we elect to only search for GCCs that are brighter than the turnover of the GCLF, $M_\mathrm{I,Vega} = -8.12$ \citep{Miller+2007}. We therefore enforce a magnitude range $-8.12 < M_\mathrm{I,Vega} < -11.5$, where the bright end roughly corresponds to the brightest known GCs. With this cut the number of GCs identified can simply be doubled to account for the uncounted fainter half of the GCLF. 

Finally, to prevent the inclusion of any remaining background galaxies we use the \texttt{Python} package \texttt{photutils} to measure the concentration index of each source (in F814W) with apertures 4 and 8 pixels in diameter \citep[as in][]{Beasley+2016b,Jones+2021}. The restrictions on this concentration index are designed to be as relaxed as possible without including large numbers of background galaxies. The locus of point sources appears at approximately $C_{4-8} = 0.45$ and we set the range as $0.2 < C_{4-8} < 0.8$. To verify that this criterion will not remove real GCs we modeled the expected concentration for a maximally extended GC. The largest GCs fit in \citet{Larsen+2001} have King profile core radii of $\sim$2.5~pc. We constructed a mock GC from a King profile with a core radius of 2.5~pc, and placed it at the nearest distance of any of the targets in our sample, 25~Mpc. This mock GC was then convolved with the average PSF in the UVIS chip (using the WFC3/UVIS PSF models in F814W provided by the Space Telescope Science Institute), then the flux was extracted within the same sized apertures as above. This gave a concentration index of $C_{4-8} = 0.73$, indicating that our concentration index cut should not exclude even the largest and nearest GCs. We also note that over all 14 targets only five potential GCCs are excluded solely on the basis of their concentration indices. We visually inspected all five and found that they were either diffraction spikes or clearly extended sources, likely background galaxies.

We inject artificial stars (point sources) into our images with true colors and magnitudes spread uniformly within our GCC selection box. These are successfully identified as GCCs approximately 90\% of the time. The main reason for the eliminations is the large color uncertainties for the faintest objects, which can result in some objects falling outside the color range used for selection. In theory objects may also scatter into our selection box, however, in practice most of the CMDs (Appendix \ref{fig:CMDs}) are so sparsely populated near the selection box that this is likely a negligible source on contaminants. We also see no change in the recovery rate towards the center of the target galaxies, where crowding might have been expected to cause issues. GCs (at least the brighter half of the GC population) appear to be sufficiently bright to prevent this crowding from playing a significant role.

\begin{figure*}
    \centering
    \includegraphics[width=\columnwidth]{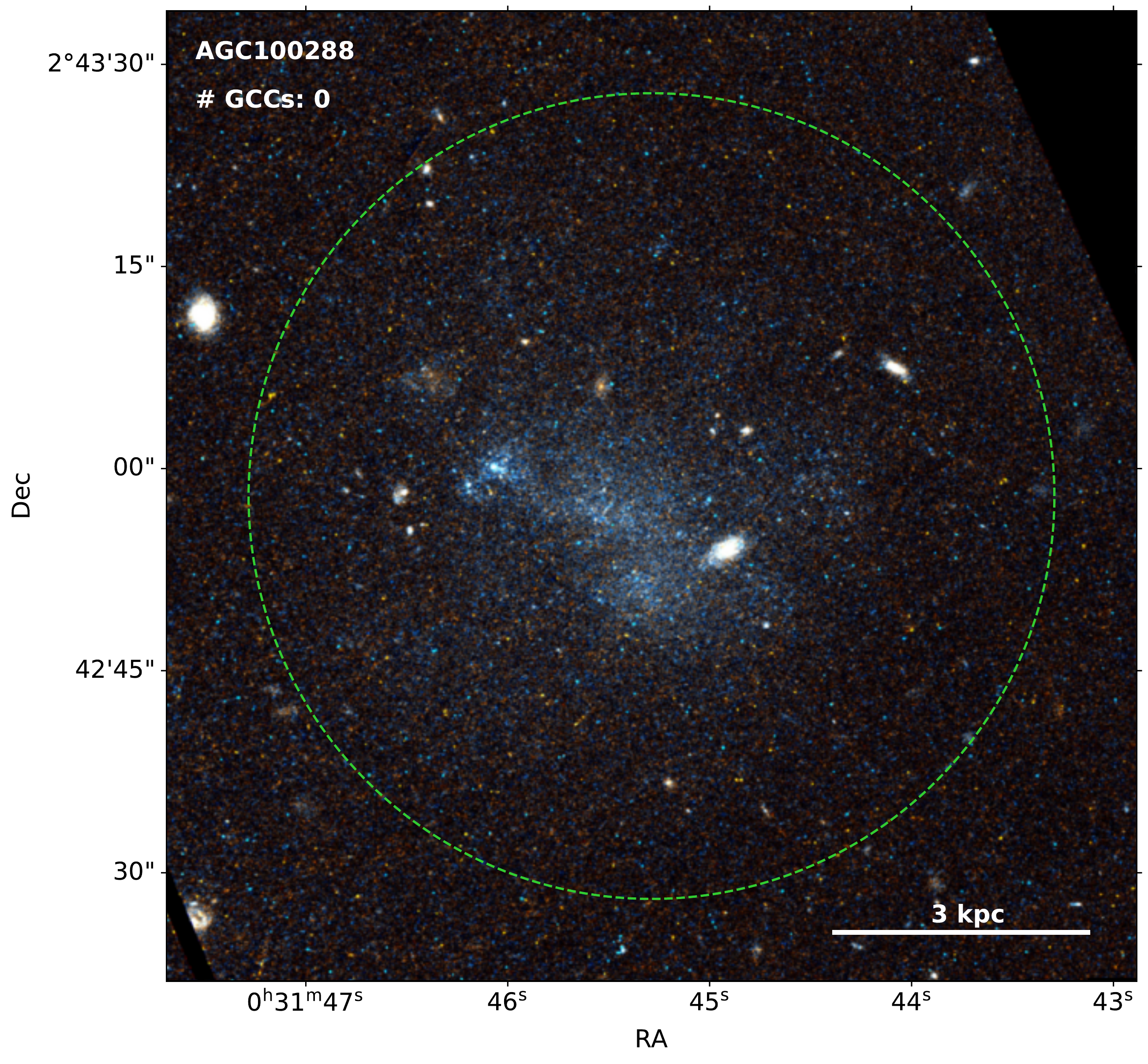}
    \includegraphics[width=\columnwidth]{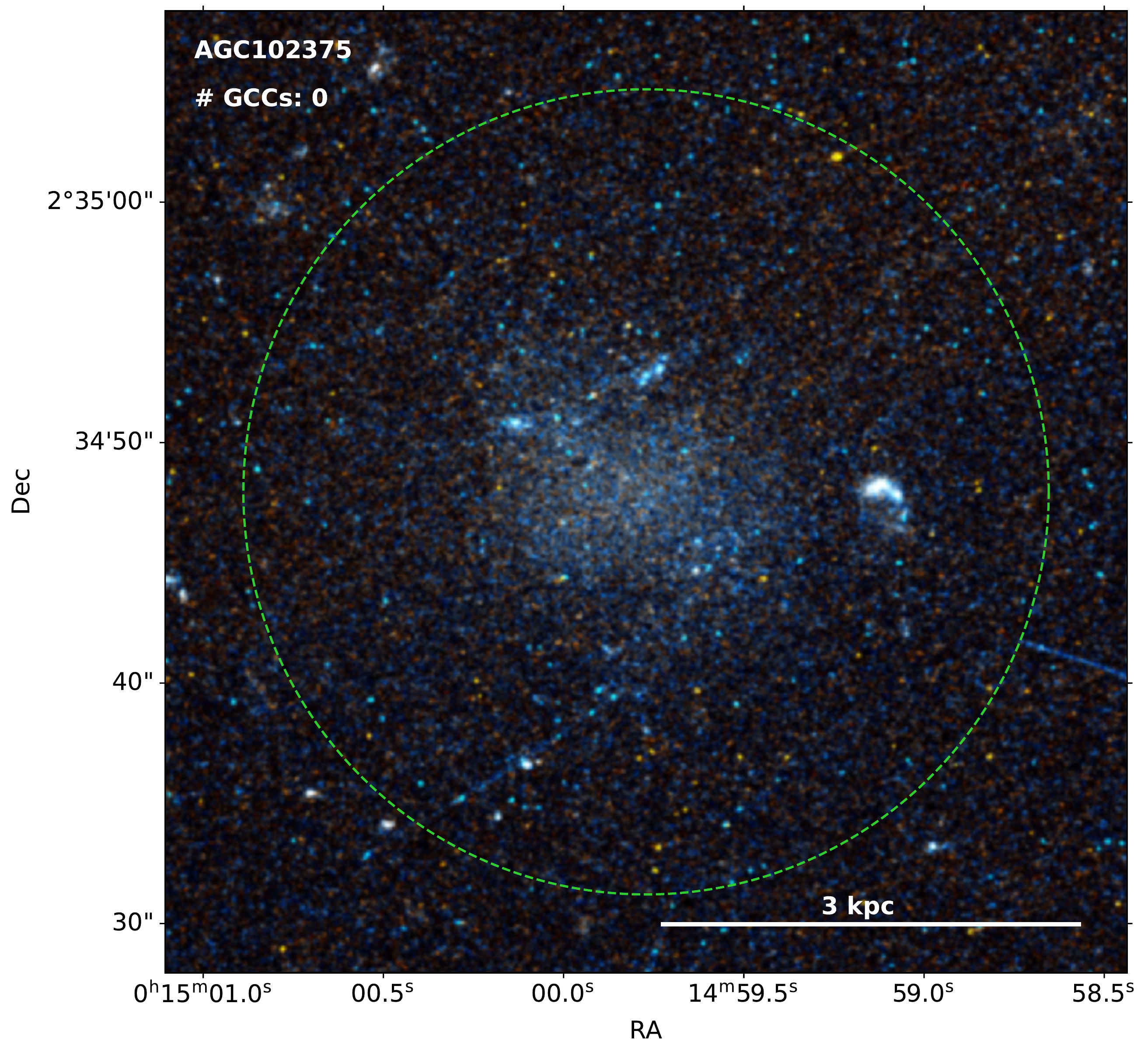}
    \includegraphics[width=\columnwidth]{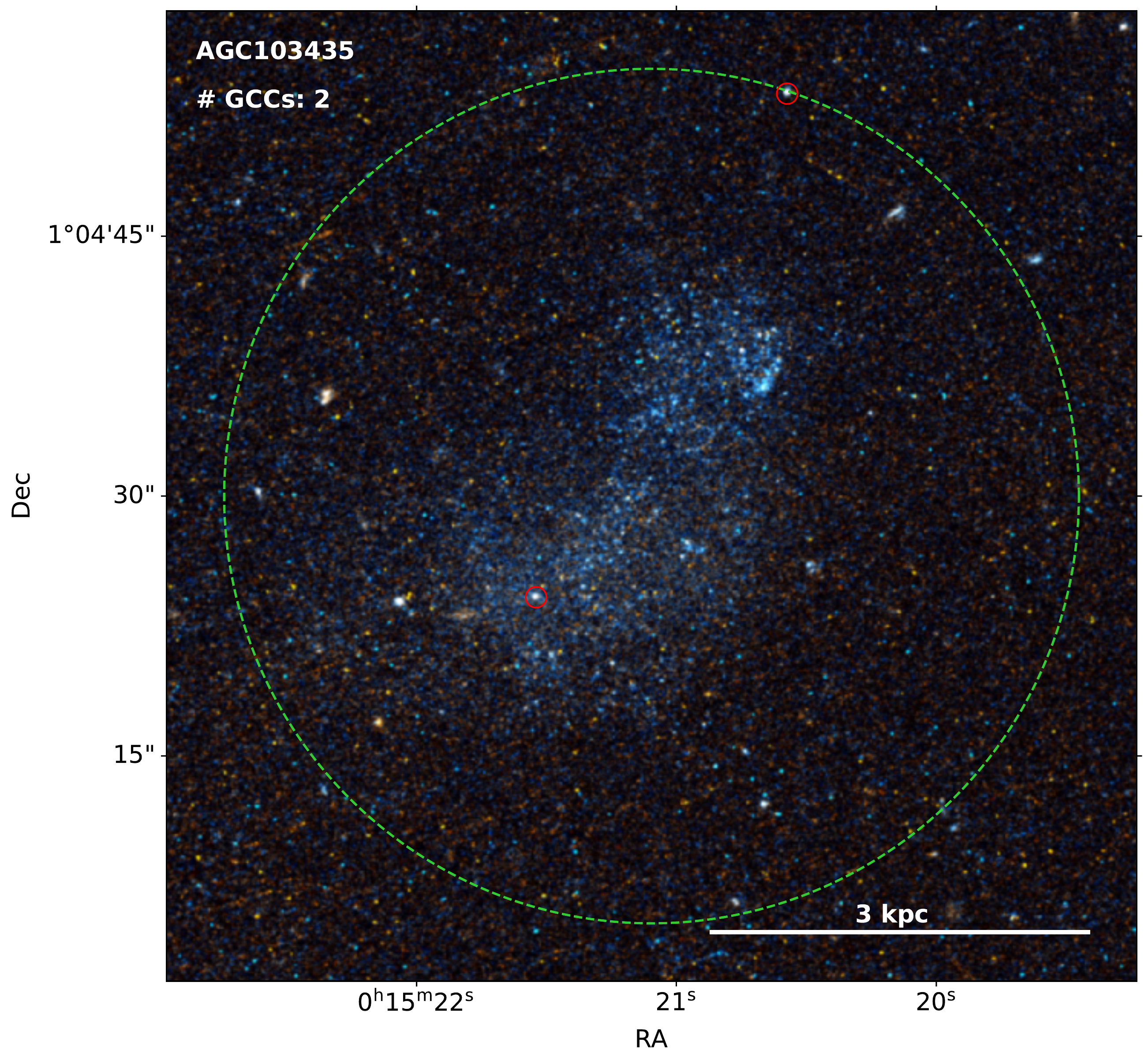}
    \includegraphics[width=\columnwidth]{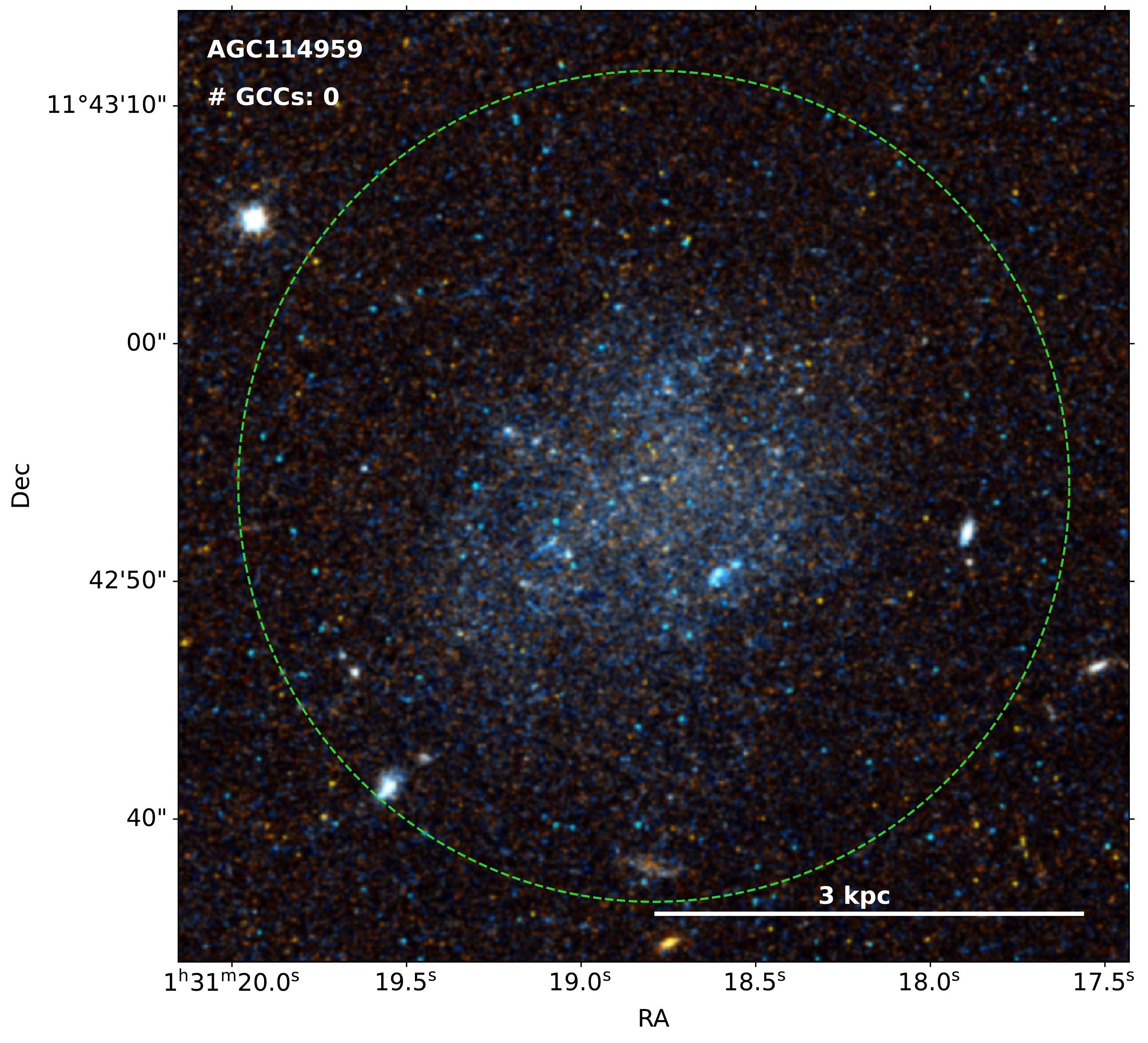}
    \caption{False color F555W+F814W WFC3 images of the target galaxies, AGC~100288, AGC~102375, AGC~103435, AGC~114959 (top-left to bottom-right). The large green dashed circles shows the region used to select GCCs, twice the half-light radius. Small red circles highlight GCCs that meet all selection criteria (\S\ref{sec:results}).}
    \label{fig:HST_images}
\end{figure*}

\begin{figure*}
    \centering
    \includegraphics[width=\columnwidth]{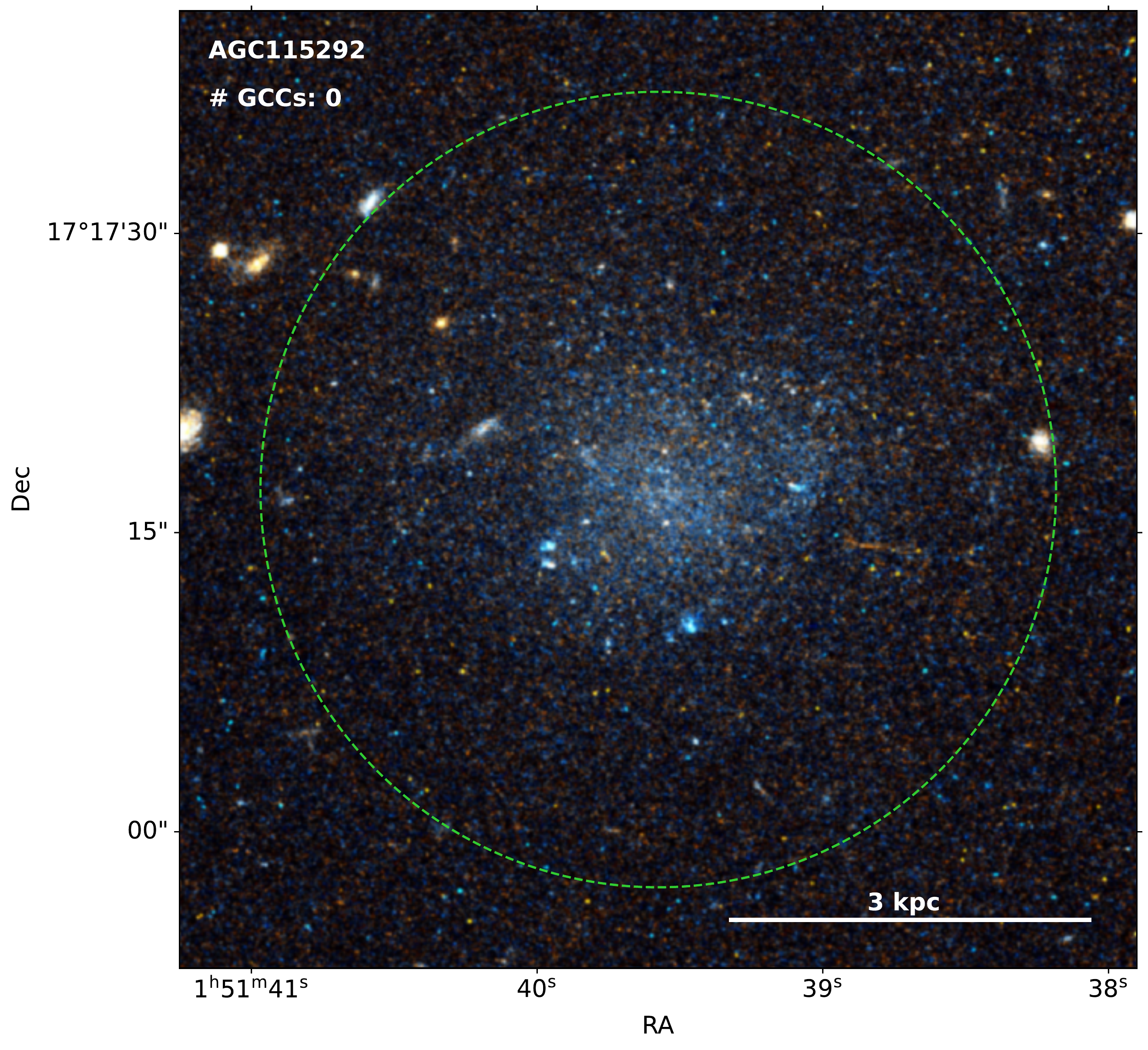}
    \includegraphics[width=\columnwidth]{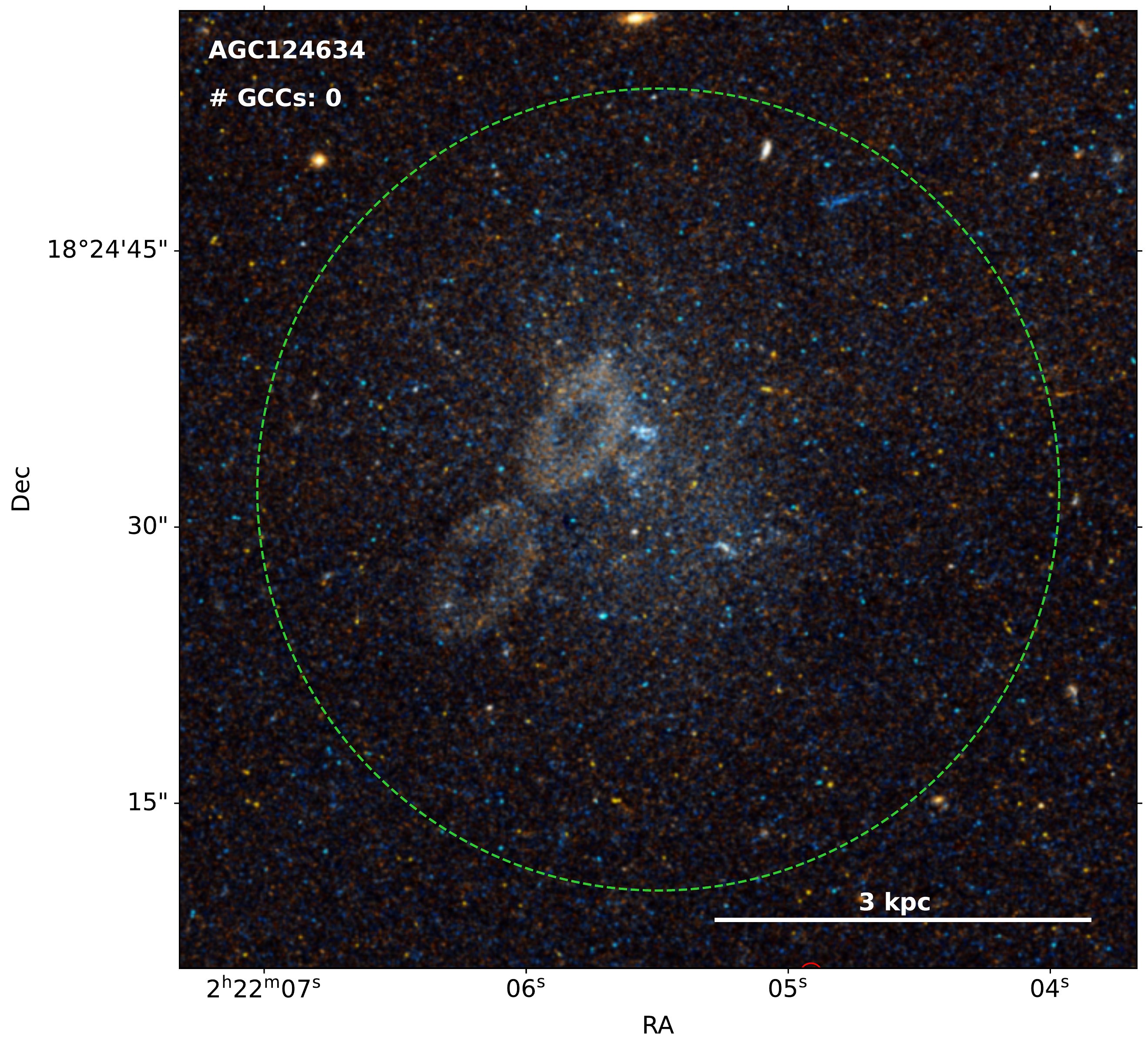}
    \includegraphics[width=\columnwidth]{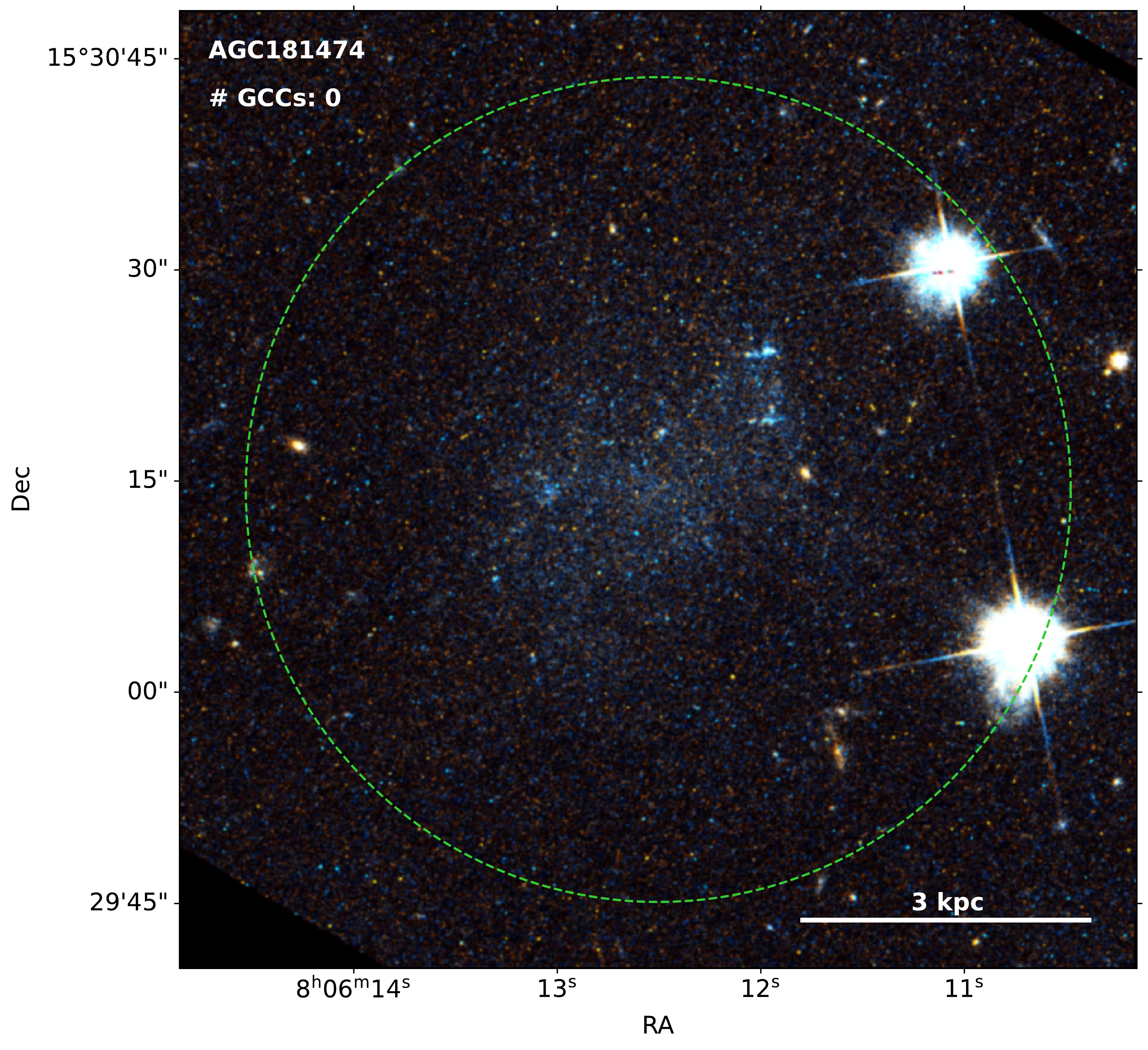}
    \includegraphics[width=\columnwidth]{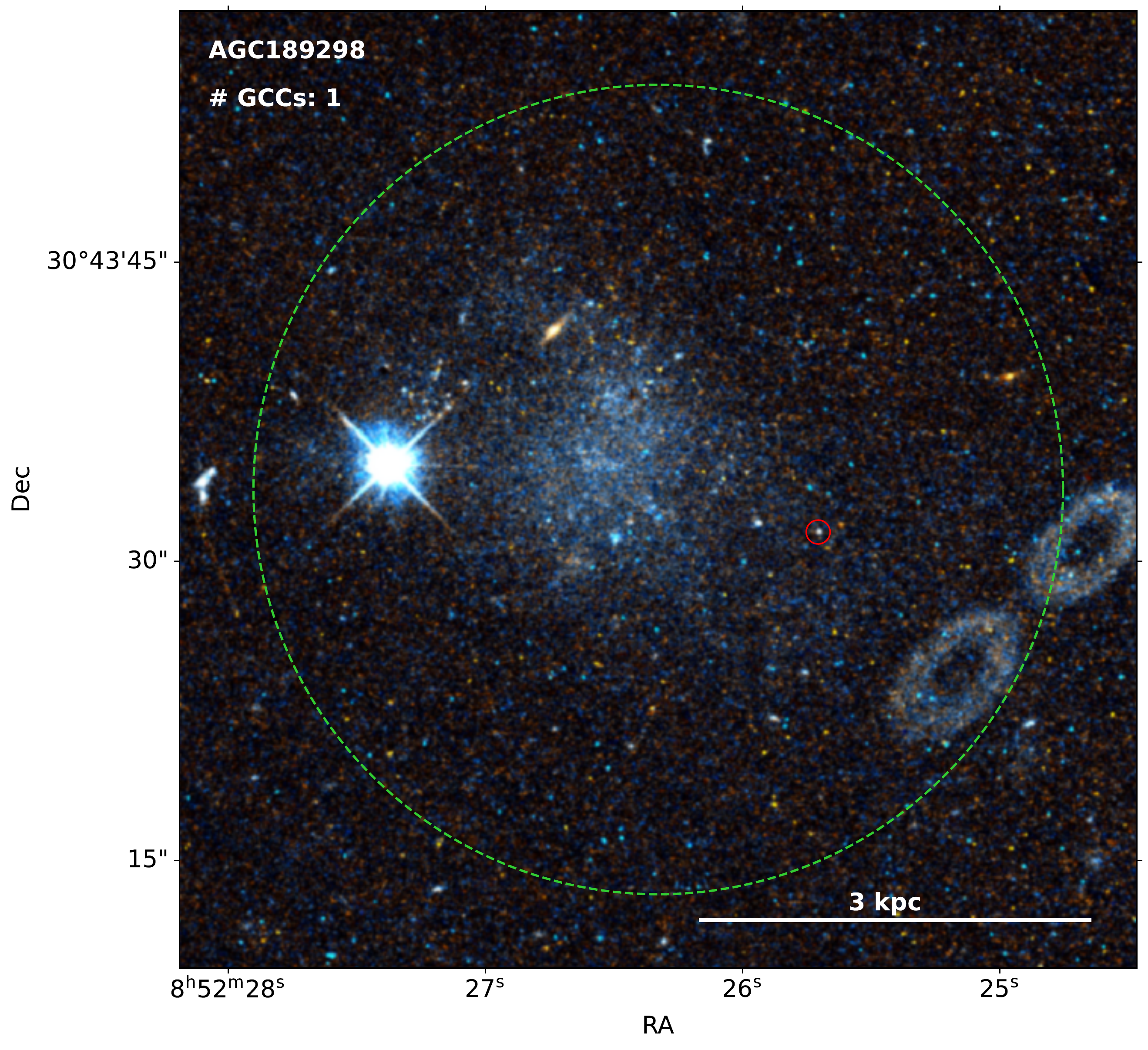}
    {\bf Figure 2 continued.} Top-left to bottom right: AGCs 115292, 124634, 181474, and 189298.
\end{figure*}

\begin{figure*}
    \centering
    \nonumber
    \includegraphics[width=\columnwidth]{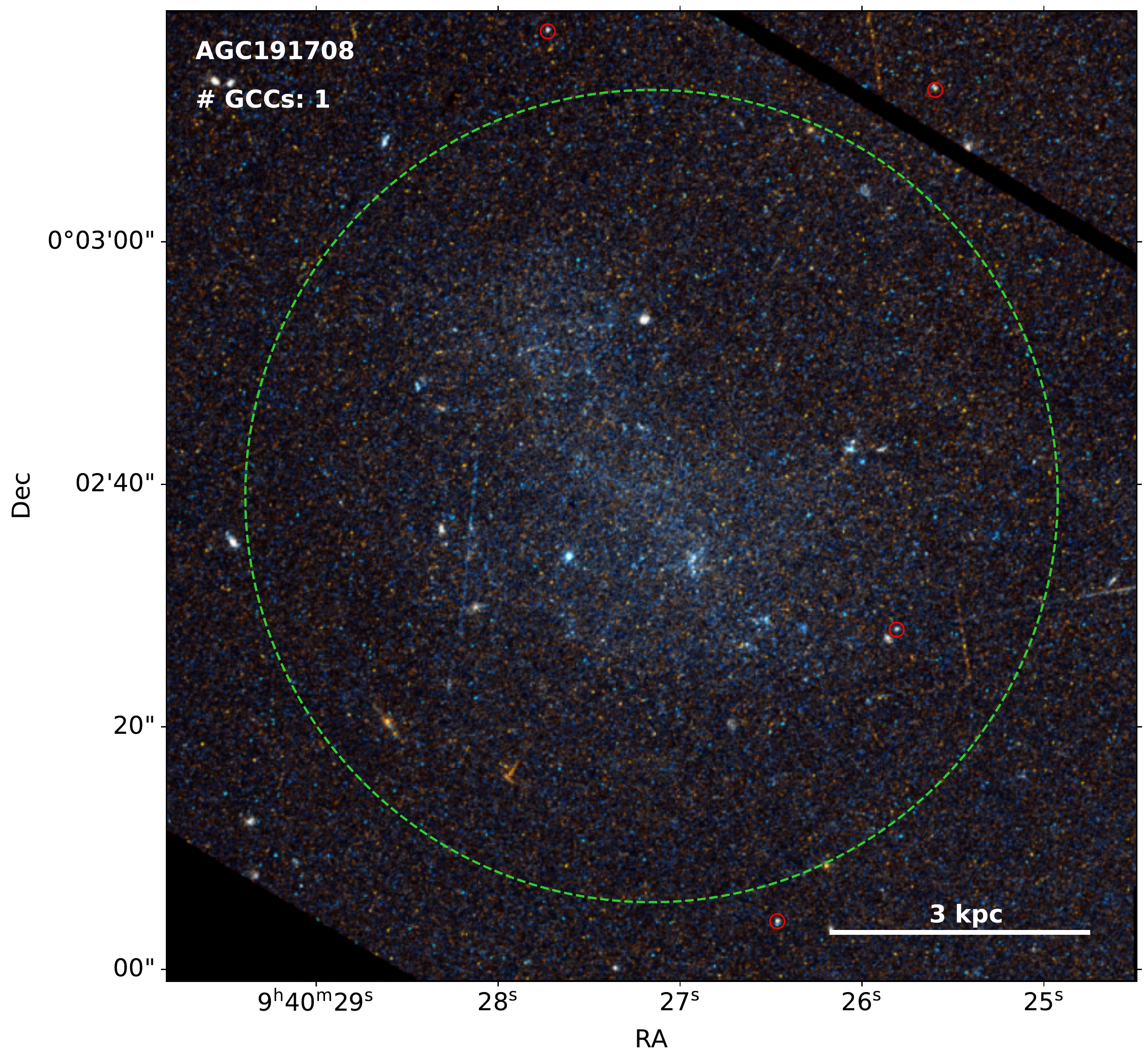}
    \includegraphics[width=\columnwidth]{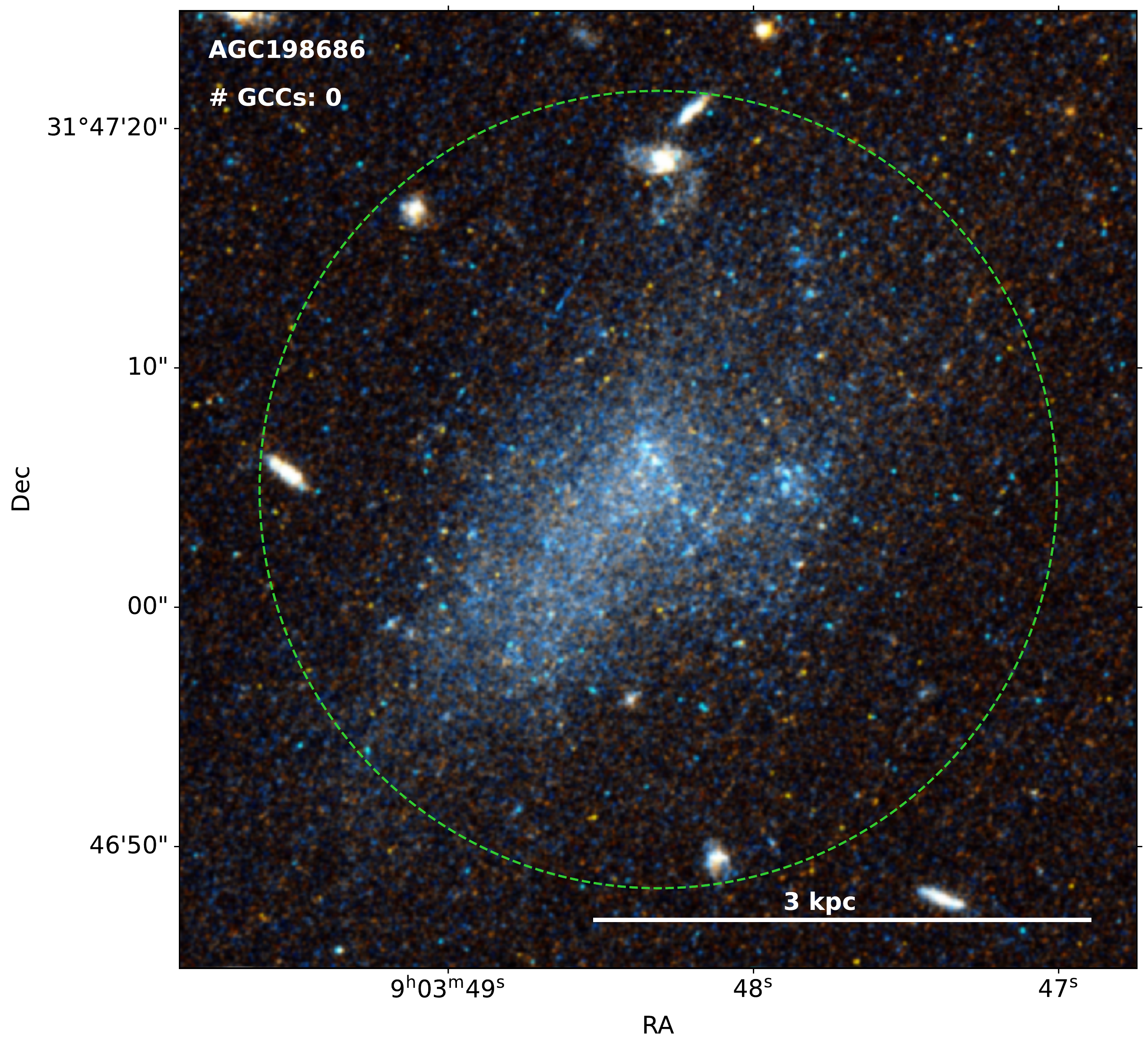}
    \includegraphics[width=\columnwidth]{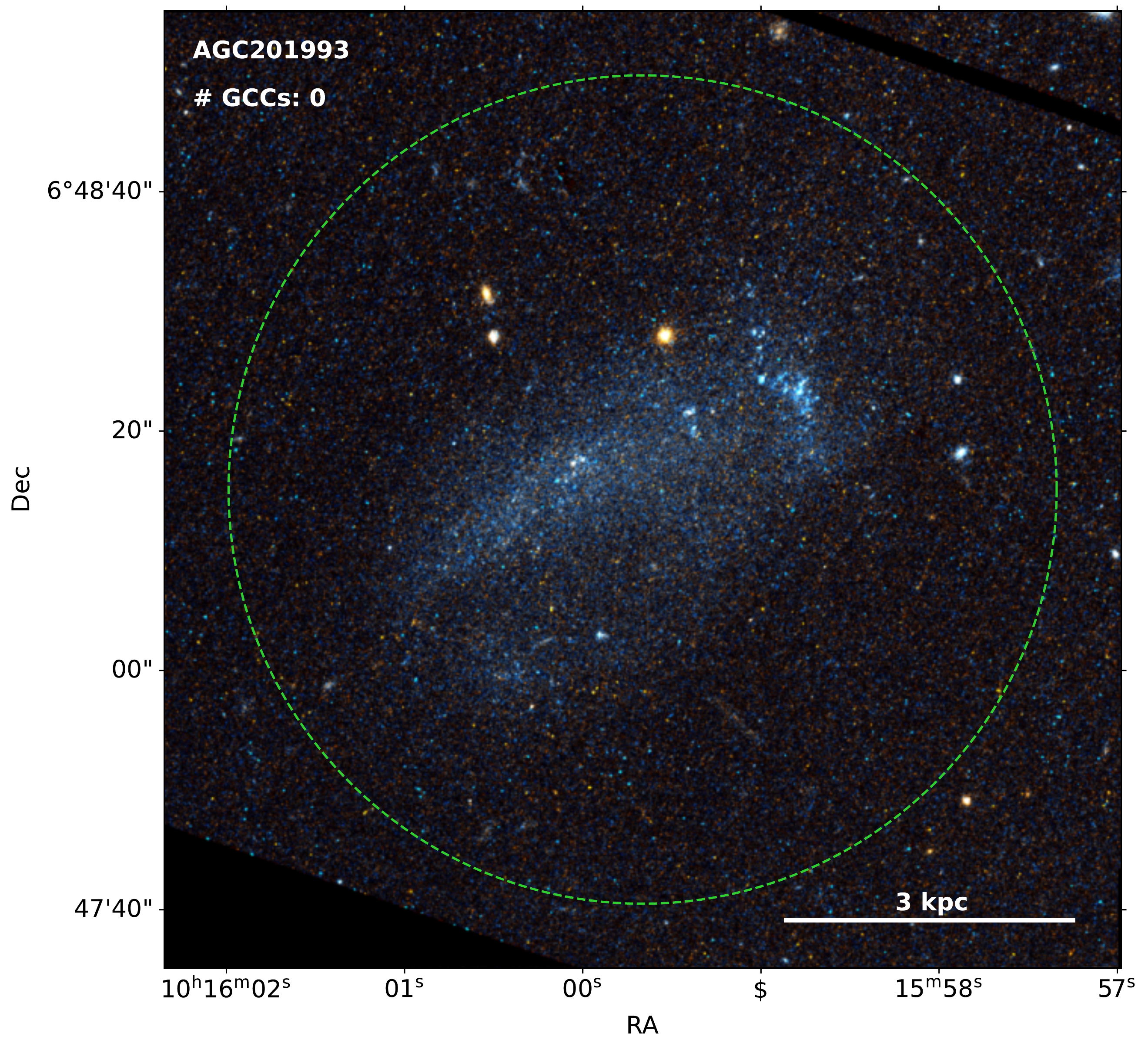}
    \includegraphics[width=\columnwidth]{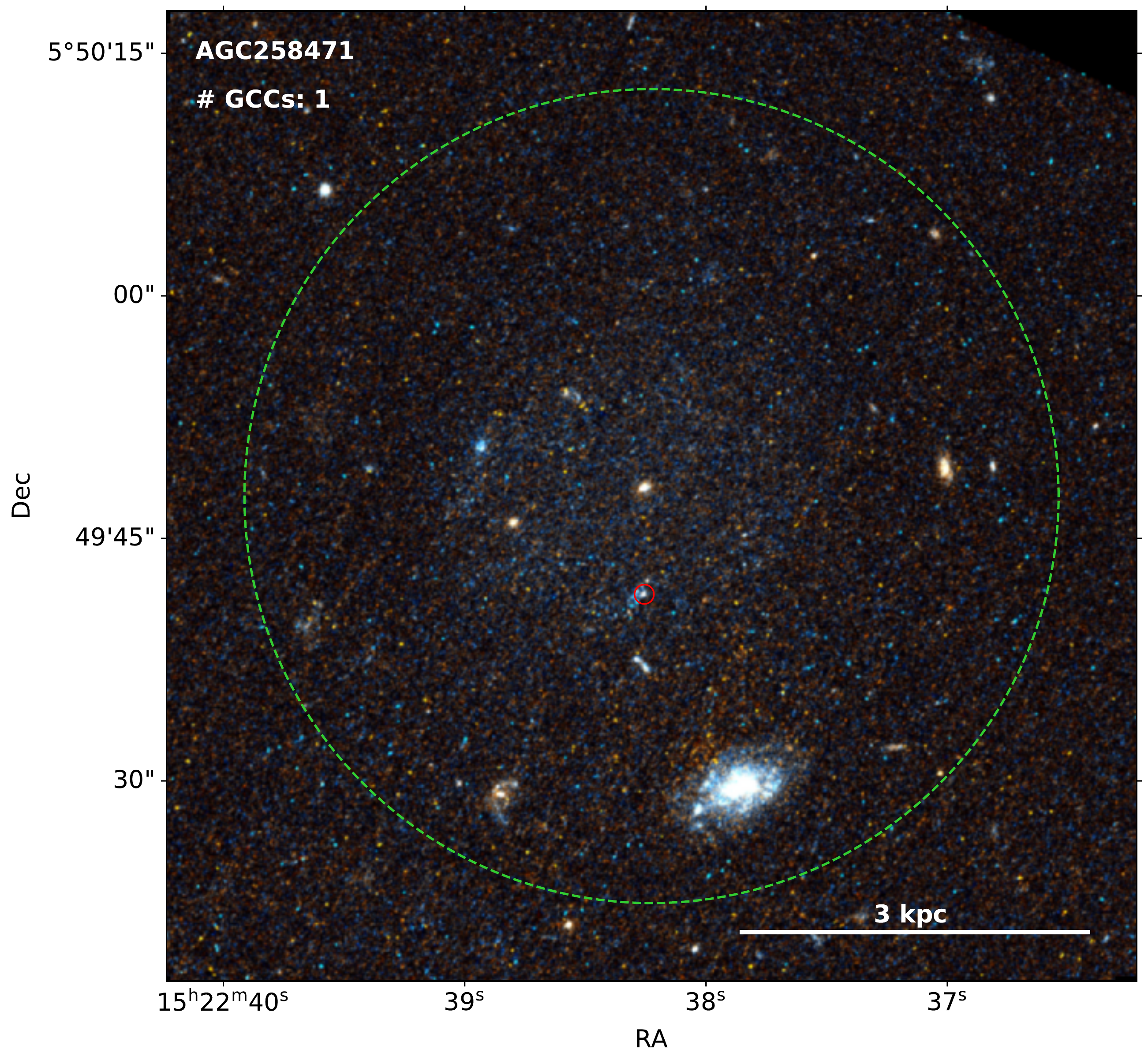}
    {\bf Figure 2 continued.} Top-left to bottom right: AGCs 191708, 198686, 201993, and 258471.
\end{figure*}

\begin{figure*}
    \centering
    \nonumber
    \includegraphics[width=\columnwidth]{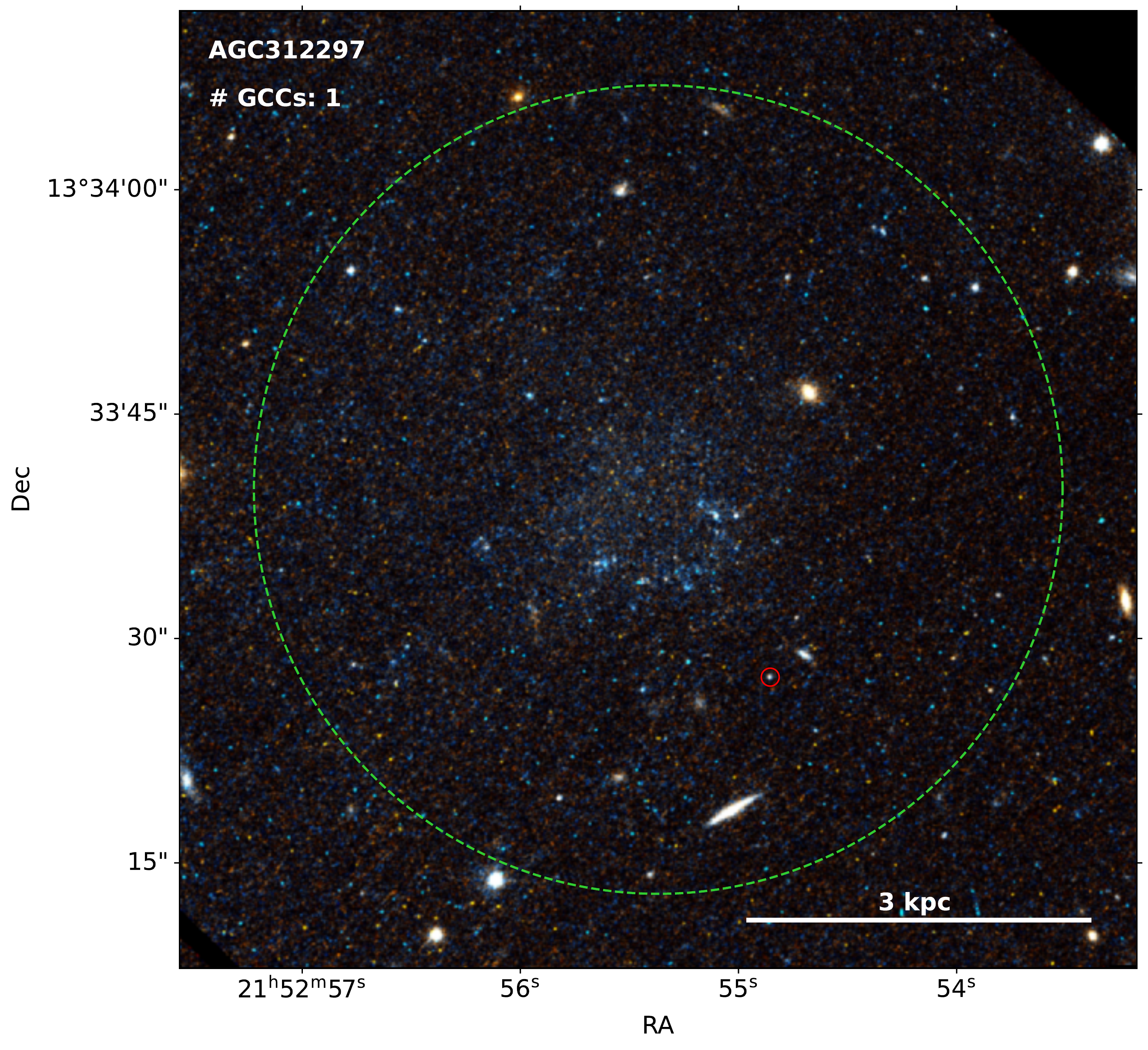}
    \includegraphics[width=\columnwidth]{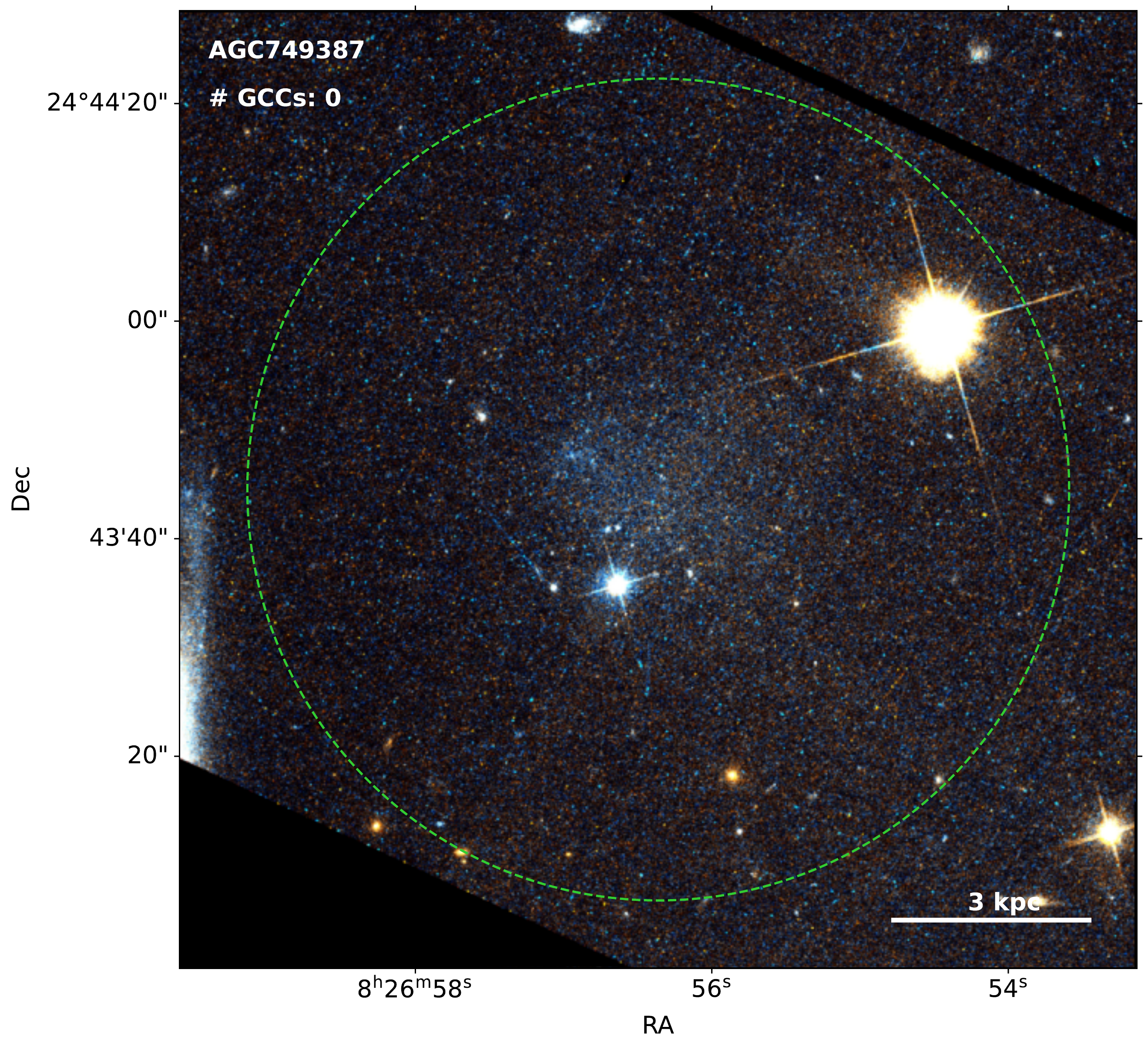}
    {\bf Figure 2 continued.} Left to right: AGCs 312297 and 749387. Note that the late-type galaxy projected next to AGC~749387 is in the background at $cz \approx 20000$~\kms \ \citep{SDSSDR6}.
\end{figure*}

Figure \ref{fig:HST_images} shows the GCCs (small red circles) identified within twice the half-light radius (dashed green circles) of each target galaxy. As is immediately apparent upon inspection, there are very few GCs in these systems. AGCs 189298, 191708, 258471, and 312297 all host a single GCC, while AGC~103435 has two \citep[all meet the UDG criteria of][]{vanDokkum+2015}. All the remaining nine systems have no GCCs.

In the case of AGC~191708 there are several potential GCCs near the edge of the area used to select objects that are likely associated with the galaxy (dashed green circle in Figure \ref{fig:HST_images}). As all our targets are highly irregular a simple selection area of twice the half-light radius might not be suitable in some cases. However, in all other cases there are no other nearby sources meeting the GCC criteria, and so this does not pose a significant issue to GCC selection. Here we simply note that AGC~191708 might be slightly anamolous relative to the rest of the sample and could potentially host several GCs if its GC system were highly spatially extended.

We estimate the false positive rate from field contaminants by counting the number of GCCs outside the encircled regions in the rest of the WFC3 FoV, and assuming that these are all false positives. After normalizing to the area of the search regions (dashed green circles, Figure \ref{fig:HST_images}) we find that the targets with the highest false positive rate are AGC~201993 and AGC~191708, with 0.8 and 0.9 false GCCs expected. However, no GCCs were identified in the former and only one in the latter. In all other cases the expectation is less than 0.25 false GCCs per target. Given these extremely low false positive rates, we elect to make no correction to the GCC counts.

It is also possible that we have eliminated some real GCs with our color criterion ($0.85 < \mathrm{V-I} < 1.5$). GCs typically follow a bimodal color distribution \citep[e.g.][]{Brodie+2006}, however, only a few percent have colors $\mathrm{V-I} < 0.85$. The CMDs in the vicinity of each target galaxy also indicate that only AGC~312297 has a significant number of objects that are just blueward of our selection box (Appendix \ref{sec:CMDs}), and these are likely young star clusters in the host galaxy, rather than genuine GCs. Therefore, we also decide to neglect this systematic correction.

\begin{figure*}
    \centering
    \includegraphics[width=0.66\textwidth]{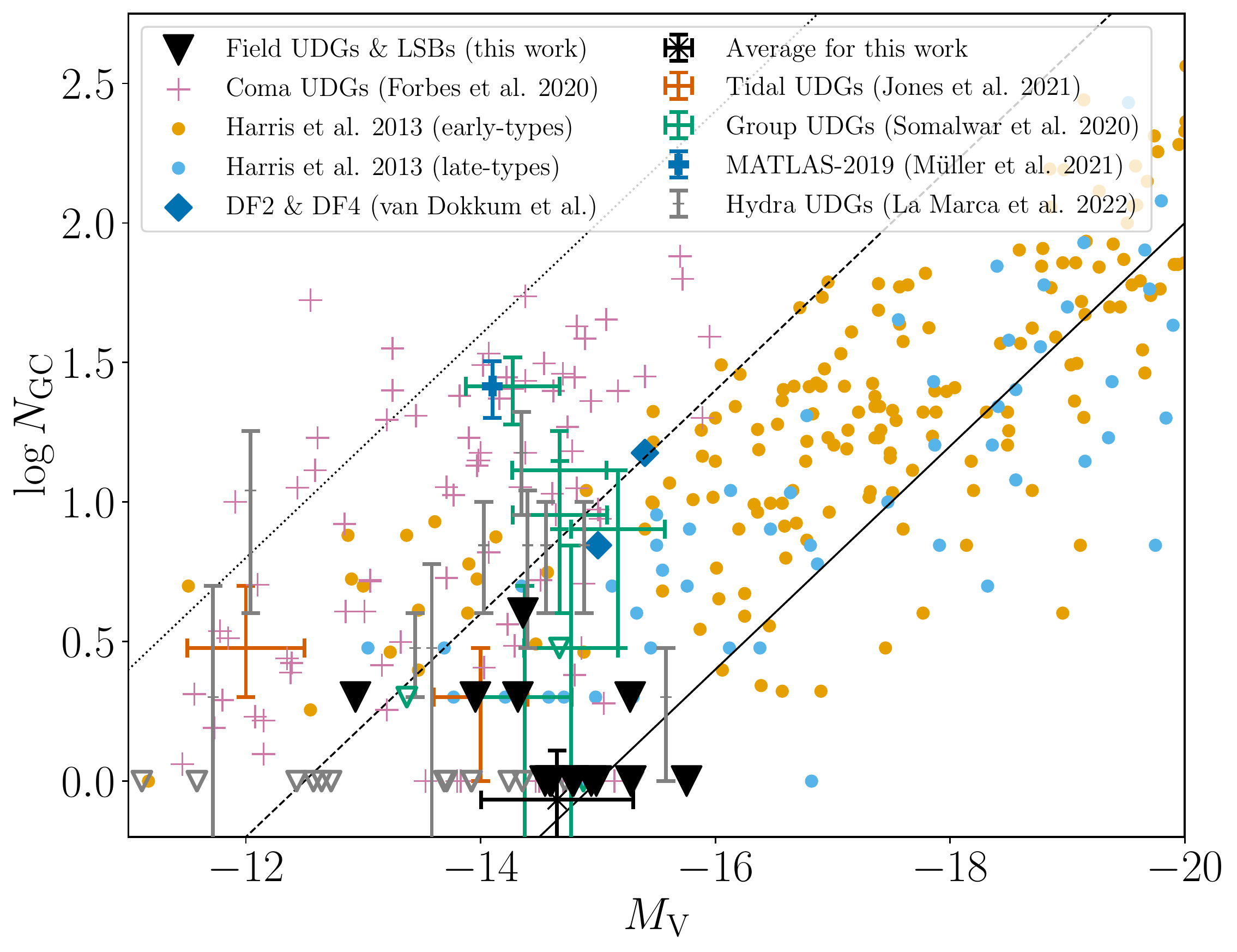}
    \caption{Globular cluster counts as a function of V-band absolute magnitude. Dwarf galaxies from \citet{Harris+2013} are shown for comparison as light orange (early-types) and light blue (late-types) circles. The GC counts \citep{Forbes+2020} of the Coma UDGs from \citet{vanDokkum+2015} and \citet{Yagi+2016} are shown with pink crosses (error bars are omitted to aid the clarity of the plot, but are typically less than 0.35~dex for $N_\mathrm{GC}$). The GC counts from this work are shown as black triangles. Those that are above zero are corrected for the missing half of the GCLF. Due to the potential for contamination from other star clusters we consider these values are upper limits on the true values of $N_\mathrm{GC}$, and thus those with no GCCs are plotted at $N_\mathrm{GC} = 1$. There are nine objects with no GCCs, however, several markers overlap (almost) entirely. The mean (corrected) number of GCs per galaxy is shown by the single set of black errorbars. UDGs in the Hydra cluster \citep{Iodice+2020,LaMarca+2022} are shown with grey errobars and upper limits (unfilled grey triangles). A small sample of UDGs in groups are also shown with green \citep{Somalwar+2020} and dark orange \citep{Jones+2021} error bars. Finally, the UDGs DF2 and DF4 \citep{vanDokkum+2018b,vanDokkum+2019} are shown with dark blue diamonds and MATLAS-2019 \citep{Muller+2021} is shown as a dark blue filled plus sign with an errorbar on $N_\mathrm{GC}$. The diagonal solid, dashed, and dotted lines correspond to GC specific frequency \citep{Harris+1981} values of 1, 10, and 100, respectively. It should be noted that the UDGs and LSB galaxies in our sample are considerably bluer than the other UDGs plotted for comparison. If plotted in terms of stellar mass, rather than $M_\mathrm{V}$, then they would shift to the left in relative to other UDGs.}
    \label{fig:NGC}
\end{figure*}

In Figure \ref{fig:NGC} we plot the number of GCs as a function of $V$-band absolute magnitude for our target sample compared to a broad sample of dwarf galaxies \citep{Harris+2013}. We double the counts for each of our targets (with $N_\mathrm{GC} > 0$) to approximately correct for the missing half of the GCLF that is fainter than our magnitude selection range. However, given the small number of GCs detected this is likely to overestimate the true number of GCs for any individual object, and we have thus plotted all values as upper limits. In the cases where no GCs were identified we set the upper limit estimate at $N_\mathrm{GC}=1$. We have also plotted the mean value for the entire sample (with the errors showing the standard deviation) which falls just below zero on the plot as the mean number of GCs per galaxy is only 0.85 after applying the factor of two correction.

Our UDG and LSB dwarfs sample appears broadly consistent with the GC counts of other dwarf galaxies, but are towards the lower limit of the luminosity range sampled by \citet{Harris+2013}. We note that galaxies with $N_\mathrm{GC}=0$ in the \citet{Harris+2013} sample are missing from this plot and it should not be used to compare to the objects in our sample where no GCs were detected. We consider these cases further in \S\ref{sec:discuss}.

Coma cluster UDGs are plotted with pink crosses. Although there is considerable scatter, on average these fall well above the $N_\mathrm{GC}$ values of the dwarf galaxies in the comparison sample. At the faintest magnitudes ($M_V \sim -12$) there may be less difference between the Coma UDGs and normal dwarf galaxies, however, the lack of objects near this magnitude in the \citet{Harris+2013} sample prevent a detailed comparison. In the magnitude range where most of our targets fall ($-16 < M_\mathrm{V} < -14$) the distribution of $N_\mathrm{GC}$ for the Coma UDGs is clearly distinct from both normal dwarfs and our gas-rich, field UDGs. However, it should be noted that the UDGs in our sample have markedly different colors from most cluster UDGs. If plotted in terms of stellar mass, then our UDGs would shift to the left and would overlap with some of the faintest cluster UDGs.

\section{Discussion}
\label{sec:discuss}

The results presented in the previous section indicate that gas-rich, field UDGs host relatively few GCs, in contrast to the many GC-rich UDGs typically found in cluster environments \citep[e.g.][]{vanDokkum+2017,Forbes+2020,Lim+2020}. In this section we discuss how these findings compare to normal dwarf galaxies and what implications this has for understanding the formation of field UDGs and their relation to UDGs in denser environments.

\subsection{Comparison to normal dwarf galaxies}
\label{sec:comp_norm_dwarfs}

Using logistic regression \citet{Eadie+2022} found that the `hurdle' stellar mass above which a galaxy is more likely to host at least one GC than to host none, is $M_\ast = 10^{6.8}$~\Msol. Based on our stellar mass estimates of our targets (Table \ref{tab:targets} and Appendix \ref{sec:stellar_masses}), all are above this threshold. From their regression model we would expect $\sim$75\% of our targets to host at least one GC. We expect to miss about half of these as our GC selection method is only sensitive to the brighter half of the GCLF. Thus, our finding that five out of 14 targets have at least one GC is broadly in line with general expectations for dwarf galaxies in this stellar mass range. While those with a non-zero number of GCs identified all fall comfortably within the scatter of the \citet{Harris+2013} sample of normal galaxies shown in Figure \ref{fig:NGC}. Therefore, in terms of their GC populations, our target sample appears to be consistent with normal dwarf galaxies.

These UDGs are also consistent with normal field dwarfs in other respects. Using a sample of 12 galaxies drawn from the same parent population, \citet{Gault+2021} found that these UDGs follow the standard \hi \ size--mass relations \citep[e.g.][]{Wang+2016}. \citet{Janowiecki+2019} also find that this population resides in an equivalent environment to other \hi-bearing galaxies of a similar mass. These results suggest that whatever the root cause of their diffuse stellar distributions, it must be an internal process rather than one governed by environment and external effects, and it likely does not strongly influence the present day distribution of the \hi \ gas.

Together all these finding point to this sub-class of UDGs being the extreme of the normal population of field dwarfs. Indeed, with the revised photometry using DECaLS imaging (\S\ref{sec:DECaLS_phot} and Figure \ref{fig:sample}), there appears to be a continuous distribution in surface brightness and effective radius from the classical dwarf regime to the most extreme objects in the sample. We also note that UDGs and LSB galaxies in the Fornax cluster show a similar trend \citep{Venhola+2017}. 

However, there are some ways, aside from their diffuse stellar distributions, that these UDGs appear to differ from normal field galaxies. It has been pointed out by multiple works \citep{Leisman+2017,Jones+2018,ManceraPina+2019,ManceraPina+2022} that they appear to be rotating more slowly than other gas-rich field galaxies, and perhaps are even DM-deficient. However, there are many uncertainties and biases that could potentially impact these findings \citep[cf.][]{He+2019}, and further investigation is required.

\citet{Kado-Fong+2022a,Kado-Fong+2022b} also recently suggested that these UDGs may have especially low star formation efficiencies (SFEs), relative to other field dwarfs, contributing to their ultra-diffuse appearance. However, we note that their control sample (which was optically-selected rather than \hi-selected) was considerably less gas-rich that the UDG sample, which itself is fairly typical of other \hi-selected galaxies in ALFALFA of similar stellar mass \citep[cf.][]{Durbala+2020}. This calls the result into question and again suggests that further work is needed to robustly contrast the star formation properties of field UDGs to normal field dwarfs. 

In summary, in terms of their GC populations, environment, and \hi \ sizes, gas-rich UDGs in the field are equivalent to other gas-rich dwarf galaxies, suggesting that they represent the extreme of a continuous distribution of surface brightness for field dwarf galaxies. However, further investigation is needed, particularly of their internal kinematics and SFEs, in order to constrain the mechanism(s) causing their diffuse stellar distributions (discussed further in \S\ref{sec:sim_comp}).

\subsection{Could field UDGs represent the progenitors of those in denser environments?}

As soon as a large number of UDGs were identified in low density environments, it was hypothesized that they could be representative of field UDGs from an earlier epoch that could have been the progenitors of present-day UDGs in clusters and groups \citep{Leisman+2017}. More recently, \citet{Junais+2022} argued that some blue UDGs in the outskirts of the Virgo cluster are being ram pressure stripped and transforming into the red UDGs typically found in cluster environments. Given that blue UDGs in the field seem to make up a significant fraction of all UDGs \citep[e.g.][]{Jones+2018,Prole+2019}, it is worth addressing the question: were the UDGs found in present-day clusters and groups once similar to the blue, gas-rich UDGs found in the field?

Old star clusters are well-suited to testing such a hypothesis as GC formation is thought to occur during the initial star formation episodes when a galaxy first forms \citep[e.g.][]{Hudson+2014}, and are therefore intricately connected to the halo mass of a galaxy. Thus, if two galaxy populations have strongly differing GC populations then their host halos must also differ, and they are unlikely to be physically related. Our findings clearly point to there being at least two pathways for forming UDGs, as the dearth of GCs in our field sample is incompatible with the GC-rich UDGs typically found in clusters. However, we note that UDGs in the Hydra cluster \citep{Iodice+2020,LaMarca+2022} appear to host far fewer GCs than those in Coma, and some of these UDGs might have had progenitors similar to the field UDGs presented in this work.

The finding that field UDGs have far fewer GCs than most cluster UDGs is in tension with the hypothesis of \citet{Junais+2022}, that blue UDGs in the outskirts of Virgo were previously field UDGs and are now transitioning to become red cluster UDGs. However, it is still possible that ram pressure stripping of field UDGs may explain a small fraction of cluster UDGs, for example, those at the lowest masses, which are less GC-rich (Figure \ref{fig:NGC}). An accounting of the GC systems of the \citet{Junais+2022} sample would help to resolve this ambiguity.

Although it is clear that gas-rich, field UDGs cannot be representative of the progenitors of most cluster UDGs, the situation with UDGs in galaxy groups is more uncertain. \citet{Somalwar+2020} measured the number of GCs in a small sample (nine) of UDGs in two galaxy groups. They found that the UDGs spanned a range of GC system richness with two objects being significantly above the distribution for normal dwarf galaxies of similar luminosity (green error bars, Figure \ref{fig:NGC}), but the remaining seven were consistent with normal dwarfs. The GC populations of the group UDGs DF2 and DF4 \citep{vanDokkum+2018b,vanDokkum+2019}, as well as MATLAS-2019 \citep{Muller+2021}, also reside at the upper limit or above those of normal dwarf galaxies. Also in a group environment, \citet{Jones+2021} found that two UDGs with evidence for tidal interactions \citep{Bennet+2018} had GCs systems consistent with normal dwarfs (dark orange error bars, Figure \ref{fig:NGC}). These results, albeit based on small samples, point to groups being an intermediate environment for UDGs, not just in the normal sense of neighboring galaxy density, but also in terms of formation pathways for diffuse galaxies. At least some UDGs in groups appear analogous to those typically found in clusters (in terms of their GC populations), while the remainder have more typical GC systems, and are presumably hosted by lower mass halos.

\citet{Jones+2021} also argued that the two group UDGs in their study were most likely formed when normal field dwarfs fell into groups and were tidally heated, resulting in a more diffuse structure \citep{Bennet+2018,Carleton+2019,Tremmel+2020}. The caveat was that, at the time, the properties of the GC systems of field UDGs were unknown. Our current finding, that gas-rich, field UDGs have GC systems that are consistent with normal dwarf galaxies, raises the possibility that these two UDGs might have been ultra-diffuse prior to falling into a group, and that the stellar streams they are adjacent to \citep{Bennet+2018} might not be indicative of the root cause of their diffuse structure. We are actively pursuing a larger sample of similar UDGs to attempt to disentangle these possibilities.

We can also consider the fate of these specific UDGs and LSB galaxies, rather than what progenitors from a past epoch they could represent. \citet{Janowiecki+2019} found that \hi-rich UDGs reside in the same environment as typical \hi-rich dwarf galaxies of similar mass. That is, they are mostly centrals in their own low-mass halos \citep{Guo+2017}. None of the 14 galaxies in our sample were matched to a known galaxy group by \citet{Jones+2020}. A visual inspection of the location of these particular galaxies reveals that those in the ALFALFA `Spring' sky are generally a few 10s of degrees away from the Virgo cluster and mostly in the vicinity of filametary structures that extend from the cluster to the east and west (also extending to higher velocities). Those in the `Fall' sky appear to be in the foreground of the Pisces-Perseus supercluster (about half way to the main structure), again in the vicinity of large-scale structures that mark the edge of a major foreground void. Thus, for these specific galaxies they will not be accreted on to a cluster for many Gyr (if ever), and are likely to remain as lone objects or perhaps join small groups.

\subsection{Comparison to simulations and models of UDG formation}
\label{sec:sim_comp}

Of the many proposed UDG formation mechanisms outlined in \S\ref{sec:intro} only those that are internal mechanisms can be invoked to explain the existence of large numbers of UDGs in the field. Currently the most favored of these are those relying on repeated episodes of star formation feedback to redistribute matter to larger radii \citep[e.g.][]{DiCintio+2017,Chan+2018} or halos in the high angular momentum tail of the naturally occurring halo spin distribution \citep{Amorisco+2016,Rong+2017}, with the angular momentum preventing an efficient collapse into a normally proportioned galaxy. Either of these models appear to be viable options for gas-rich, field UDGs. 

\citet{Trujillo-Gomez+2022} and \citet{Danieli+2022} also suggest that the formation of GCs themselves might be responsible for feedback that causes the diffuse structure of UDGs. However, for these models to explain our gas-rich, field UDGs they would likely need to host more GCs than we have identified, unless the majority have been lost. 

Attempts have been made to measure the gas kinematics of these UDGs \citep{ManceraPina+2019,ManceraPina+2020,ManceraPina+2022} as a means to probe their specific angular momenta. However, the poor resolution of most of the currently available data is problematic for drawing robust conclusions. In addition, even with accurate modeling of the kinematics of the baryons, there is no guarantee that this maps directly to the global DM halo angular momentum, which is not observable. 

In the case of the star formation feedback models, much higher angular resolution kinematic data would be needed to directly identify the cored DM profiles that these models predict. Another means to test this model would be to obtain high temporal resolution star formation histories. However, all the objects in this sample are too distant to do this with HST, but it may be possible for some objects with the James Webb Space Telescope. Given these challenges, it is important to ask what constraints can be drawn from the finding that these galaxies host few GCs.

When discussing the GC populations of Coma cluster UDGs, \citet{Saifollahi+2022} highlighted that their high values of $N_\mathrm{GC}/M_\ast$ is an argument against forming UDGs purely by redistributing stellar mass to large radii, regardless of the exact mechanism for doing so. The formation pathway must include a mechanism that either increases the number of GCs or suppresses the expected stellar mass, or both. In the case of gas-rich, field UDGs, where the GC counts are equivalent to normal dwarf galaxies, this argument works in reverse. The correct explanation for their diffuse structure likely involves an almost purely re-distributive process that does not significantly impact either GC formation or the overall build-up of stellar mass. 

Complicating this somewhat, \citet{Gault+2021} found that the \hi \ gas is not in a more extended distribution than is typical (for the total \hi \ mass), therefore the redistribution process must only significantly affect stars (and potentially DM), not gas. As \hi \ is generally much more spatially extended than the stars in most gas-rich galaxies, this may, for example, be possible with a cored DM halo, if the core is sufficiently compact as to not strongly influence the distribution of gas in the galaxy outskirts.

Recent hydrodynamical simulation results from IllustrisTNG \citep{Nelson+2019a} indicate that late-type LSB galaxies form in higher mass halos than higher surface brightness galaxies of the same stellar mass \citep{Perez-Montano+2022}. Expanding on this, \citet{Benavides+2022} found that UDGs in the TNG50 simulation \citep{Pillepich+2019,Nelson+2019} are also skewed toward higher halo masses than normal dwarf galaxies, regardless of environment. The stellar mass estimates for the UDGs in our sample are typically $\log M_\ast/\mathrm{M_\odot} \approx 7.5$, which, for a field UDG, corresponds to a halo mass of $\log M_{200}/\mathrm{M_\odot} \approx 10.5$, according to \citet{Benavides+2022}.\footnote{We note that this mass is at the lower extreme of their UDGs in their sample.} \citet{Zaritsky+2022} measured the linear relationship between the number of GCs and total galaxy mass, finding that on average there is 1 GC per $(2.9 \pm 0.3) \times 10^9$~\Msol \ of total galaxy mass. If we use the halo mass estimate above as the total mass for our UDGs, then we would expect the UDGs in our sample to typically host $10.9 \pm 1.1$ GCs (note that this ignores the uncertainty in the halo mass estimate). Thus, the relative lack of GCs that we find is in clear tension with the expectation from TNG50 \citep{Benavides+2022}. Given that the GC counts that we find appear to be compatible with normal dwarf galaxies, this suggests that these UDGs are in fact not hosted in DM halos that are more massive than those of other galaxies of similar stellar mass. If there are many unidentified field UDGs (analogous to DGSAT~I, \S\ref{sec:DGSAT}), then the findings of \citep{Benavides+2022} may offer an explanation, but this is not consistent with gas-rich, field UDGs.

\citet{Wright+2021} reported an alternative formation pathway for field UDGs in the Romulus25 simulation \citep{Tremmel+2017}. In this case UDGs were formed primarily through low-mass galaxy mergers that redistributed star formation more towards galaxy outskirts. This produced simulated field UDGs that have typical star formation rates and colors for field galaxies of similar stellar masses. The abundance of field UDGs in this model also matches quite well with estimates from \citet{Jones+2018}. As the relation between halo mass and $N_\mathrm{GC}$ is linear \citep{Zaritsky+2022}, this scenario should also result in UDGs with GC systems comparable to normal dwarfs of similar masses.

\subsection{Comparison to DGSAT~I}
\label{sec:DGSAT}

The UDG DGSAT~I \citep{Martinez-Delgado+2016} is one of the few known quenched, field UDGs. It is located in the vicinity of the Pisces-Perseus supercluster at a distance of approximately 80~Mpc. A recent study of its GC system \citep{Janssens+2022} produced the opposite finding to our sample of gas-rich, field UDGs. DGSAT~I hosts a rich ($N_\mathrm{GC} = 12 \pm 2$) and compact GC system, much more in line with UDGs in clusters. 

It is possible that DGSAT~I-like objects are the progenitors of cluster UDGs, however, it is a peculiar object even among UDGs, and such a conclusion would be premature without a larger sample of similar quenched field UDGs being identified first. What is clear is that DGSAT~I did not form via the same pathway as the gas-rich UDGs we consider in this work. As discussed by \citet{Janssens+2022}, the most straightforward explanation may be that DGSAT~I is a backsplash object that formed via the same mechanism as cluster UDGs.

\section{Conclusions}
\label{sec:conclude}

We have imaged 14 gas-rich, field UDGs and LSB galaxies with HST WFC3, and selected GCCs based on color, magnitude, and concentration index. We find strikingly few candidates, in stark contrast to the GC-rich UDGs typically found in galaxy clusters. Nine of our 14 targets have no GCCs brighter than the turnover magnitude of the dwarf galaxy GCLF ($M_\mathrm{I,Vega} = -8.12$).

These low GC counts are consistent with expectations for normal dwarf galaxies in a similar stellar mass range ($\log M_\ast/\mathrm{M_\odot} \sim 7.5$), and suggest that the formation process driving the diffuse structure of these galaxies is primarily a re-distributive process that moves stars to larger radii, without significantly impacting the formation of GCs, the long term build-up of stellar mass, or the present day distribution of neutral gas.

Given the small number of GCs in these field UDGs and LSB galaxies, they cannot represent the progenitors of red, GC-rich UDGs in clusters, which presumably formed in higher mass halos. However, they may be the progenitors of some UDGs in group environments, which seem to exhibit a broad range of GC richness.

\begin{acknowledgments}
We thank the anonymous referee for their comments which helped to improve this letter. This work is based on observations made with the NASA/ESA Hubble Space Telescope, obtained at the Space Telescope Science Institute, which is operated by the Association of Universities for Research in Astronomy, Inc., under NASA contract NAS5-26555.  These observations are associated with program \# HST-SNAP-16758.  Support for program \# HST-SNAP-16758 was provided by NASA through a grant from the Space Telescope Science Institute.
This work used images from the Dark Energy Camera Legacy Survey (DECaLS; Proposal ID 2014B-0404; PIs: David Schlegel and Arjun Dey). Full acknowledgment at \url{https://www.legacysurvey.org/acknowledgment/}.
DJS acknowledges support from NSF grants AST-1821967 and 1813708.
KS acknowledges support from the Natural Sciences and Engineering Research Council of Canada (NSERC).
AK acknowledges financial support from the grant (SEV-2017-0709) funded by MCIN/AEI/ 10.13039/501100011033 and from the grant POSTDOC\_21\_00845 funded by the Economic Transformation, Industry, Knowledge and Universities Council of the Regional Government of Andalusia.
LL acknowledges support from NSF grant AST-2045371.
\end{acknowledgments}

%

\vspace{5mm}
\facilities{HST(WFC3), Blanco}


\software{\href{http://americano.dolphinsim.com/dolphot/}{\texttt{DOLPHOT}} \citep{Dolphin2000}, \href{https://www.astropy.org/index.html}{\texttt{astropy}} \citep{astropy2013,astropy2018}, \href{https://photutils.readthedocs.io/en/stable/}{\texttt{Photutils}} \citep{photutils}, \href{https://reproject.readthedocs.io/en/stable/}{\texttt{reproject}} \citep{reproject}, \href{https://sites.google.com/cfa.harvard.edu/saoimageds9}{\texttt{DS9}} \citep{DS9}, 
\href{https://github.com/ConnorStoneAstro/AutoProf}{\texttt{AutoProf}} \citep{Stone+2021}, \href{https://github.com/
danjampro/DeepScan}{\texttt{DeepScan}} \citep{Prole+2018}
          }



\appendix

\section{Color-Magnitude Diagrams}
\label{sec:CMDs}

\begin{figure*}
    \centering
    \includegraphics[width=0.24\textwidth]{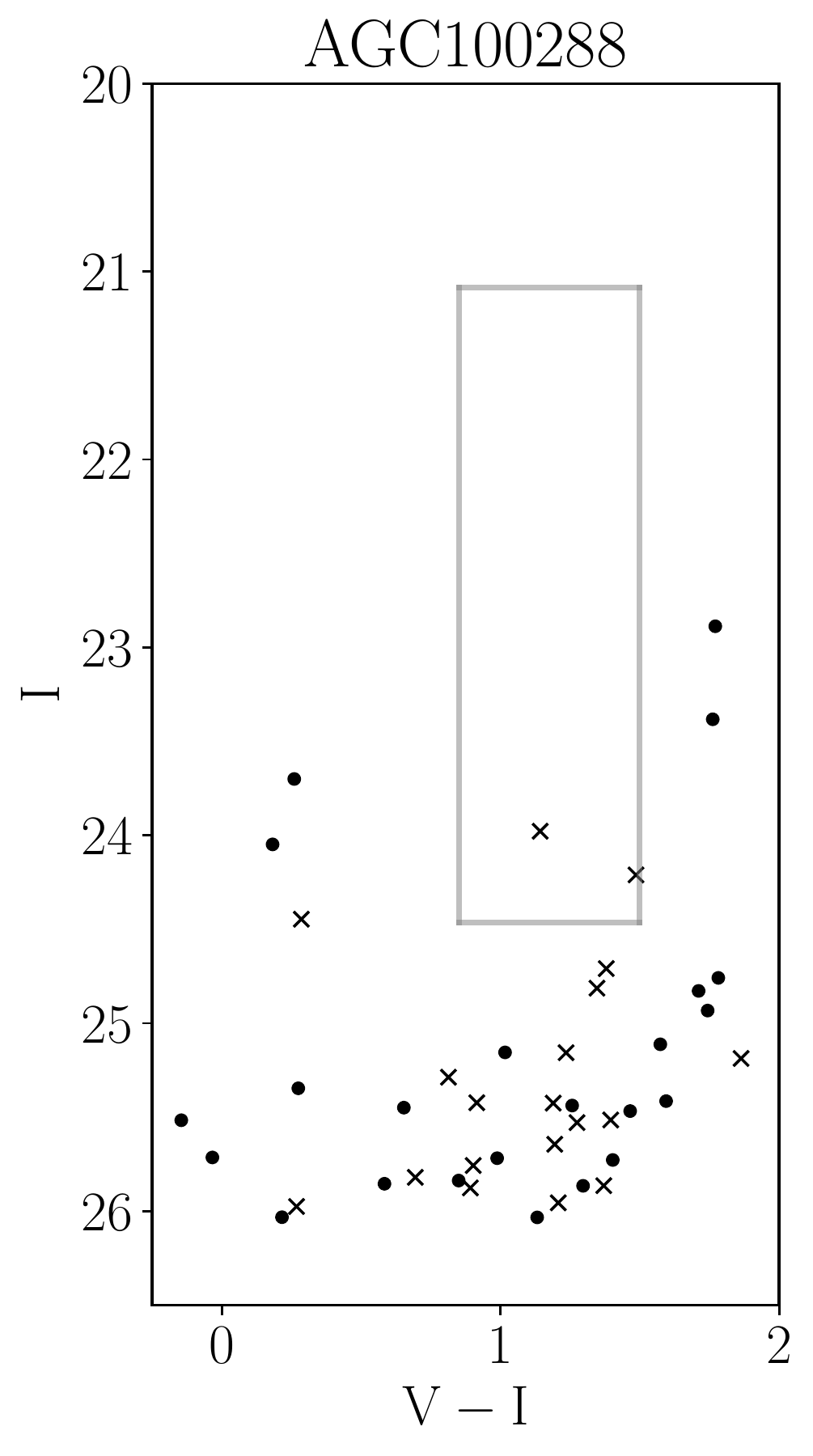}
    \includegraphics[width=0.24\textwidth]{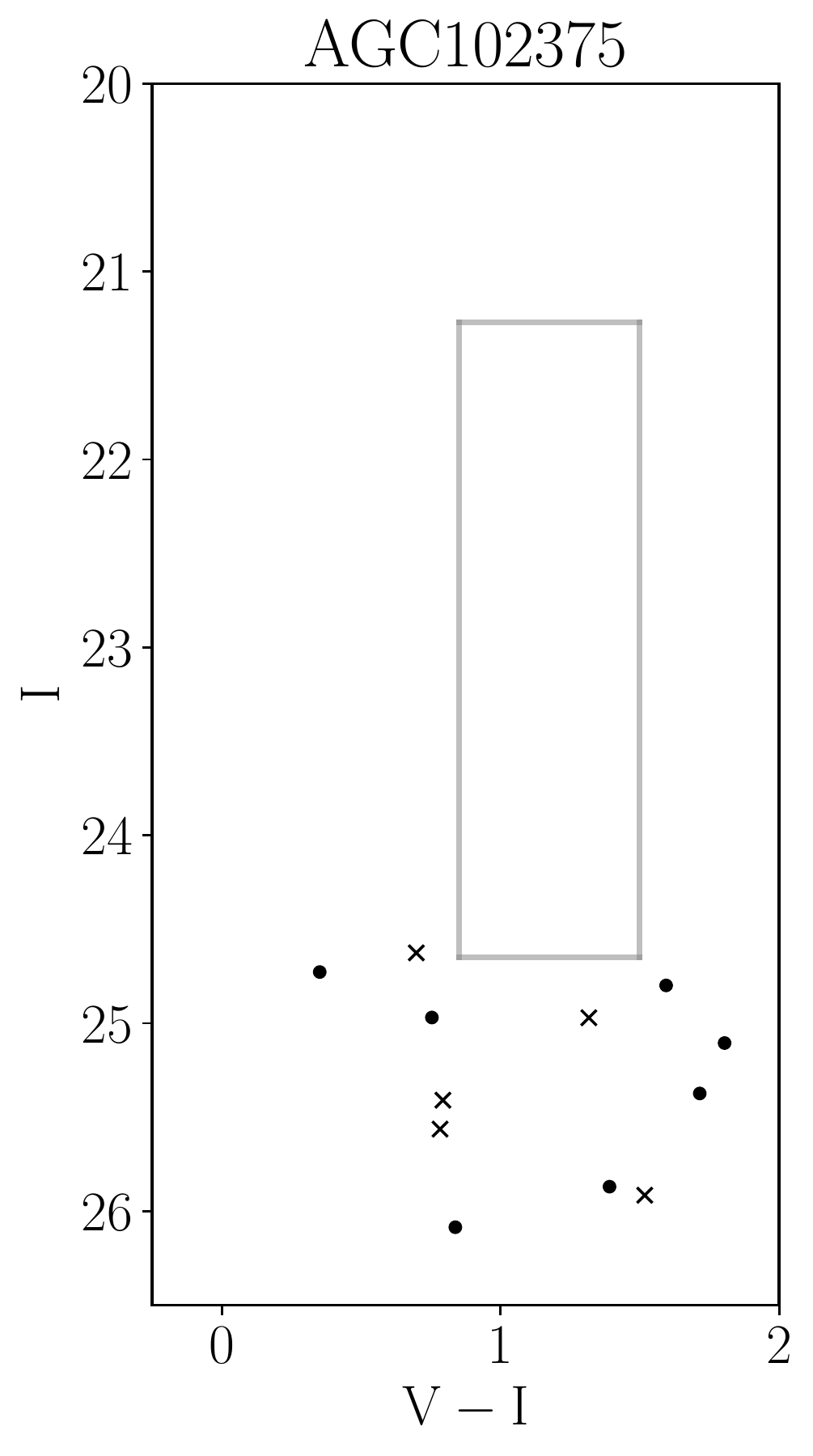}
    \includegraphics[width=0.24\textwidth]{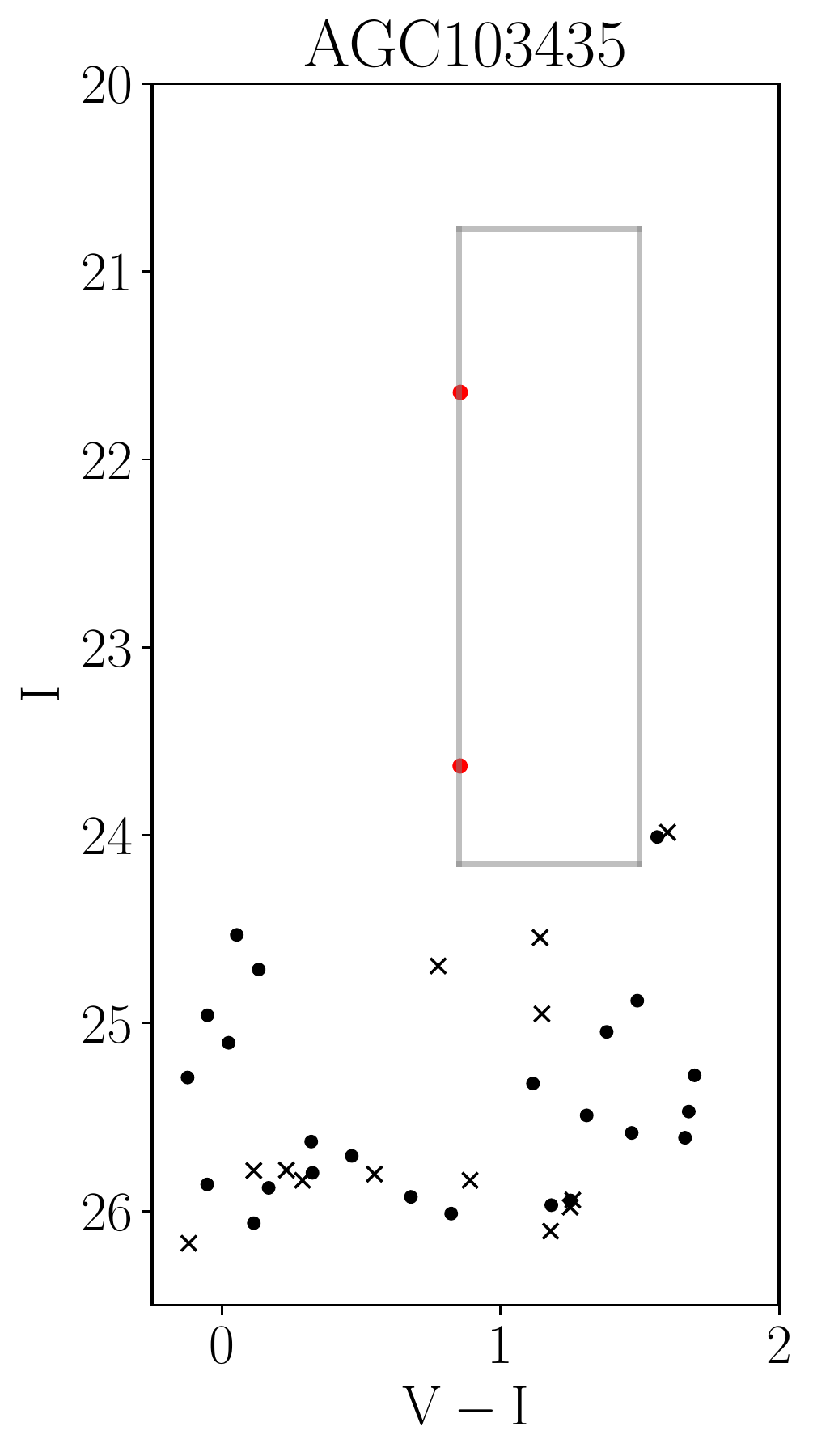}
    \includegraphics[width=0.24\textwidth]{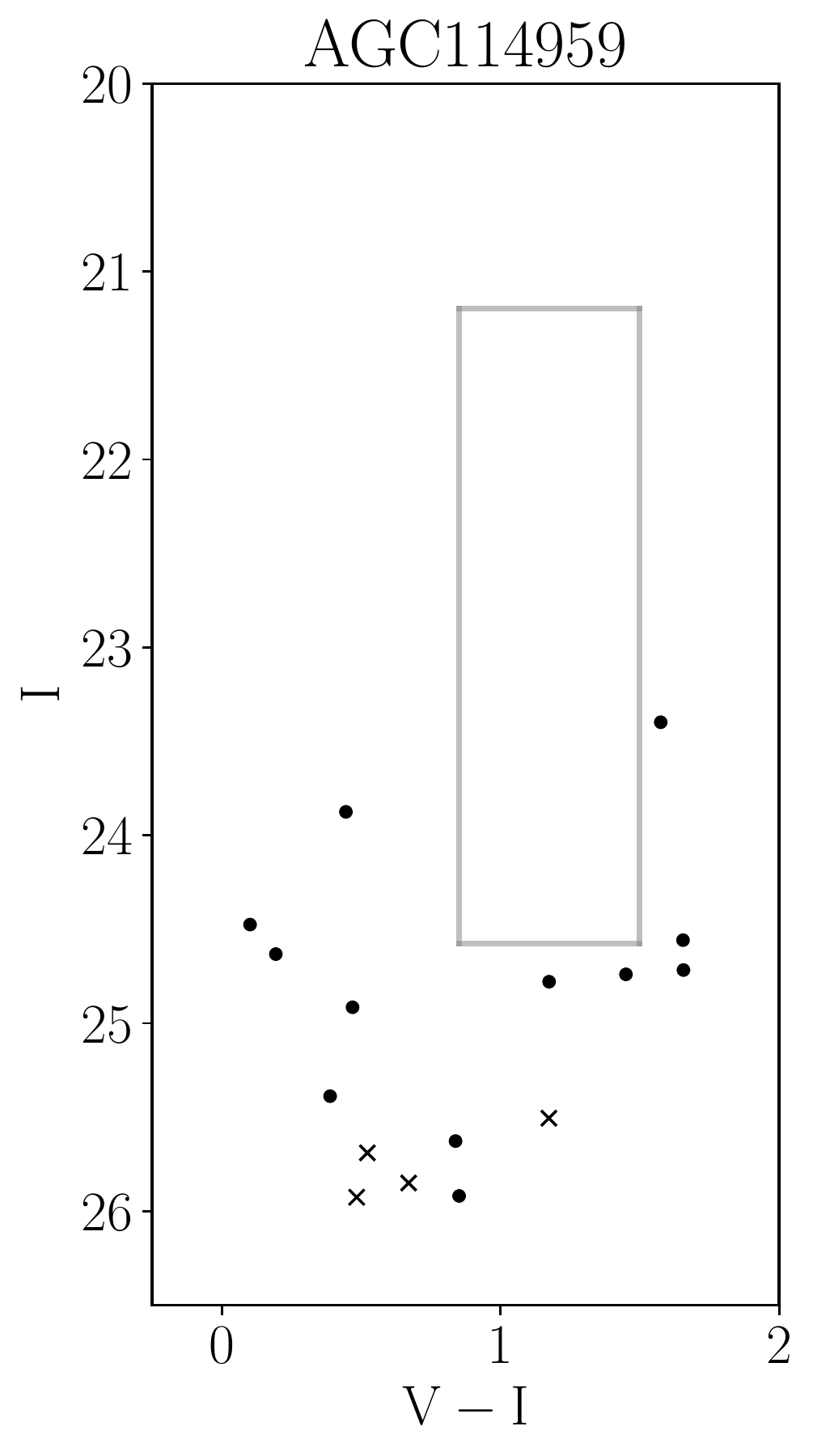}
    \includegraphics[width=0.24\textwidth]{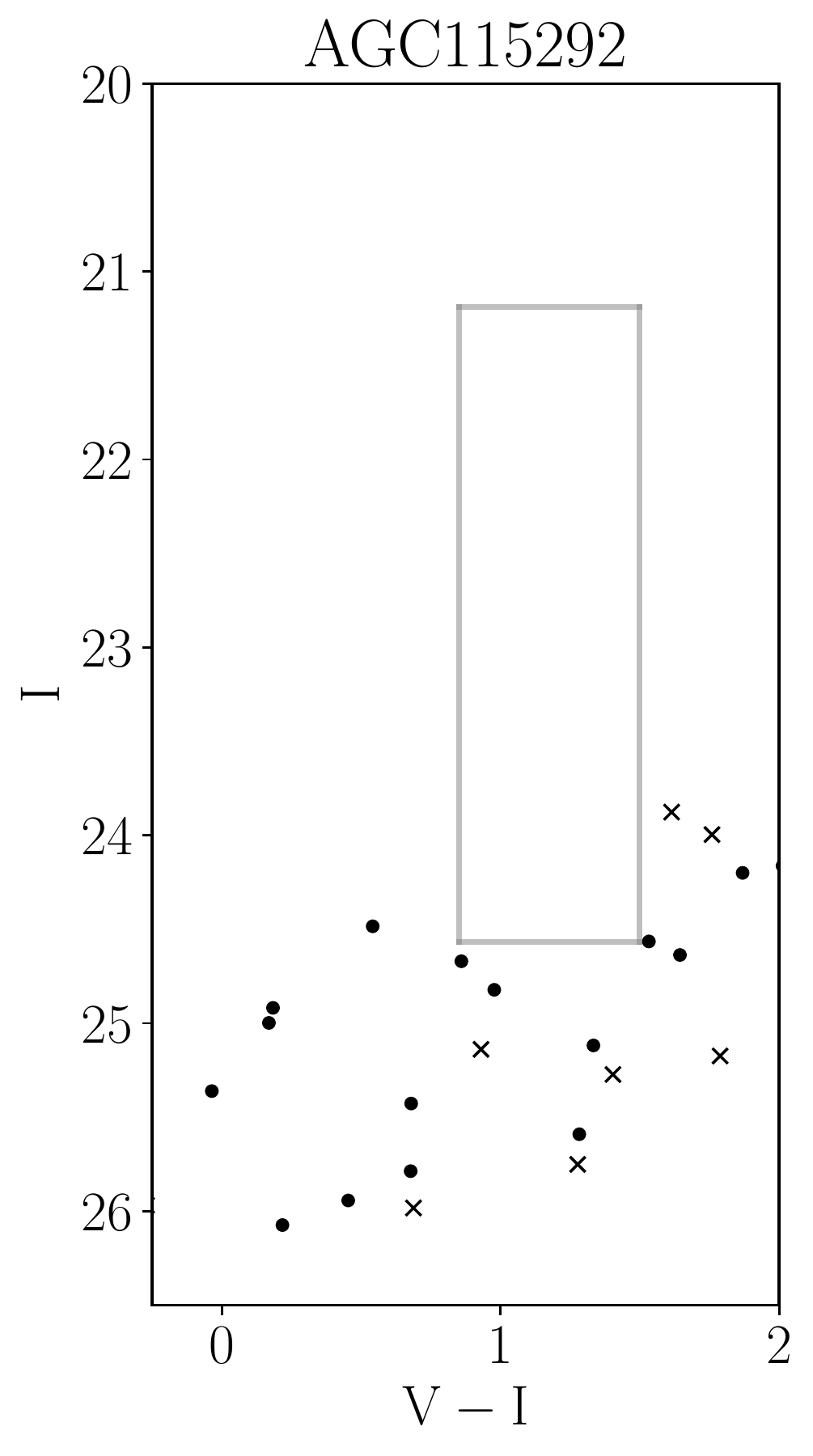}
    \includegraphics[width=0.24\textwidth]{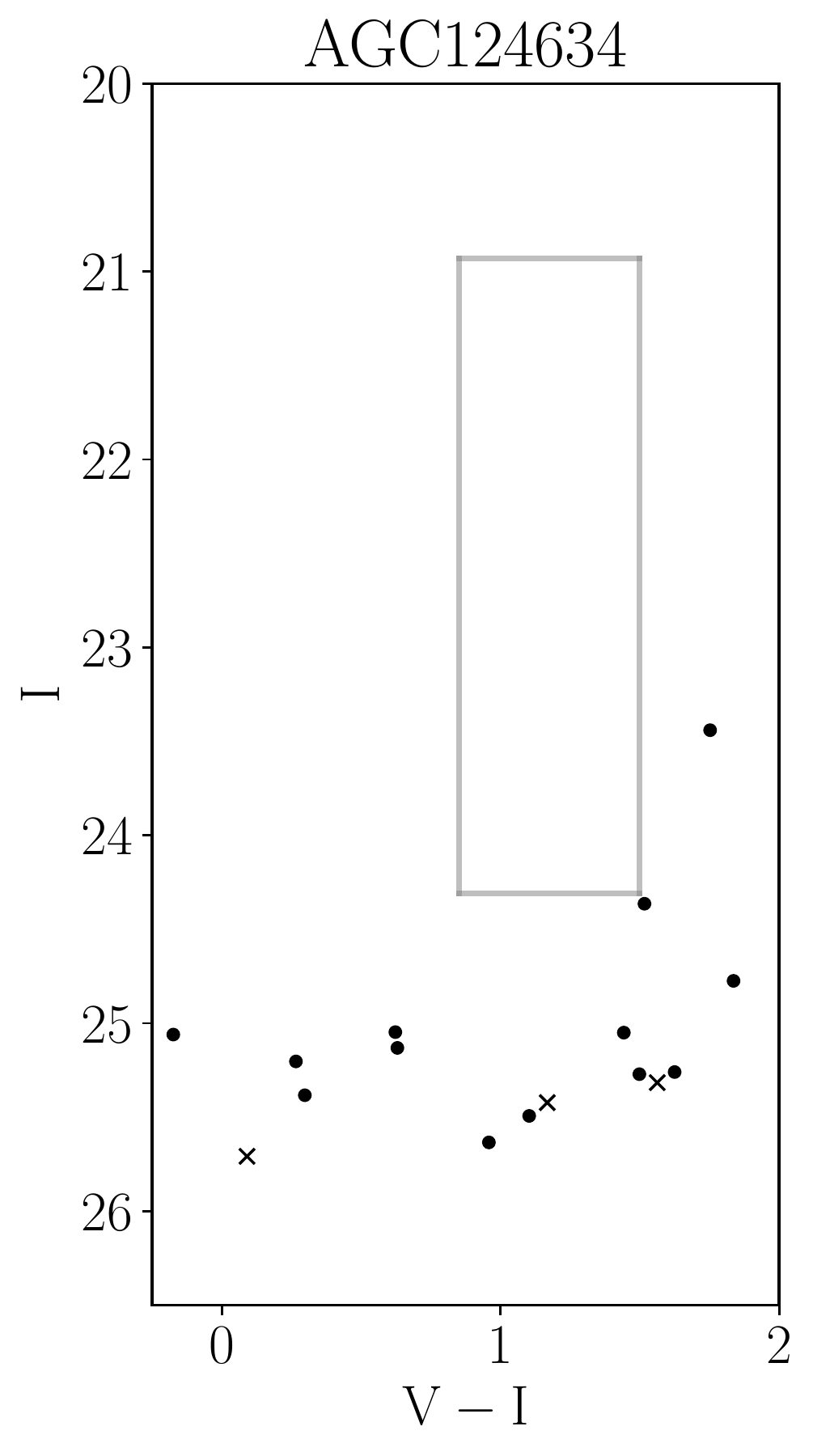}
    \includegraphics[width=0.24\textwidth]{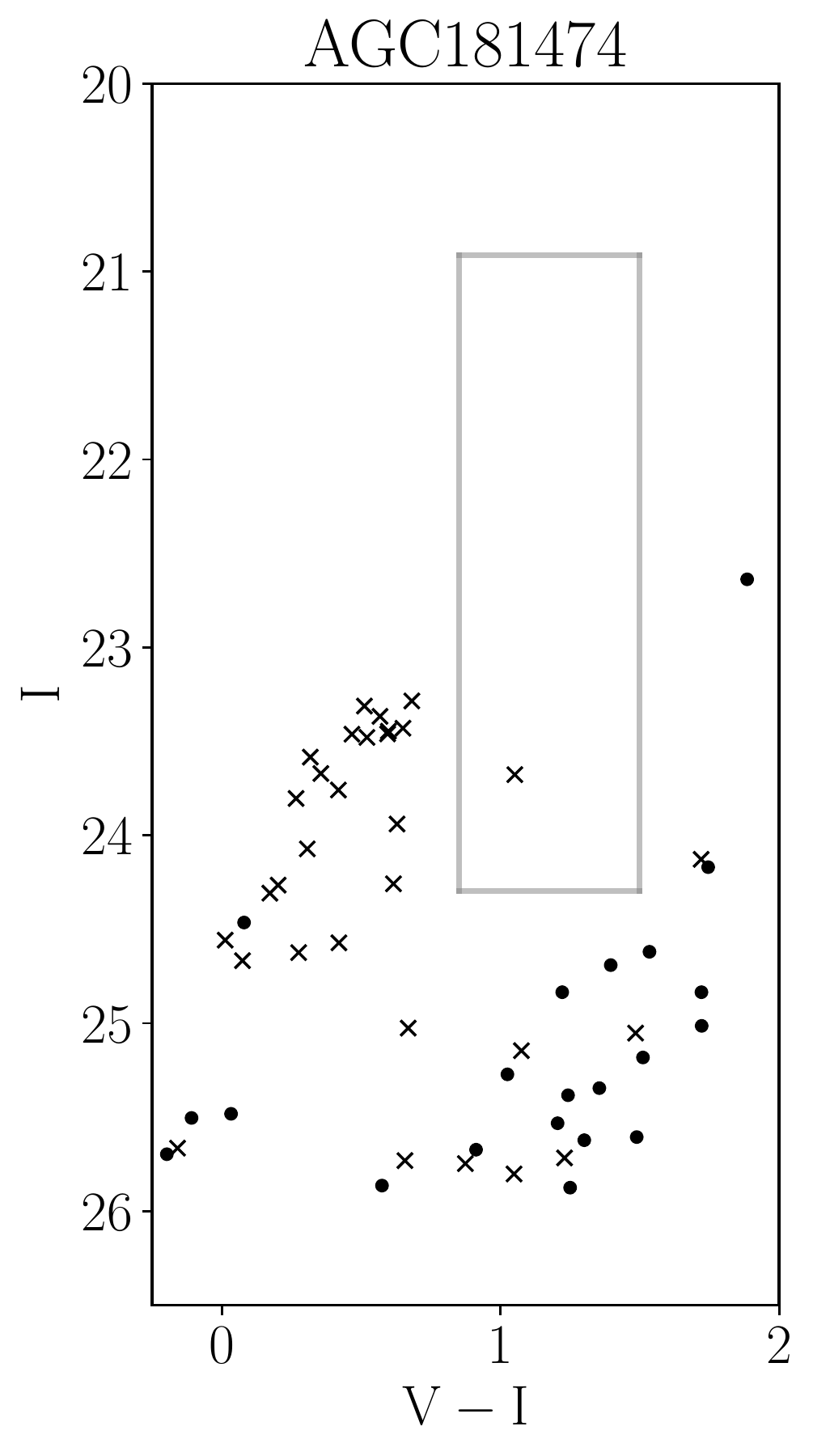}
    \includegraphics[width=0.24\textwidth]{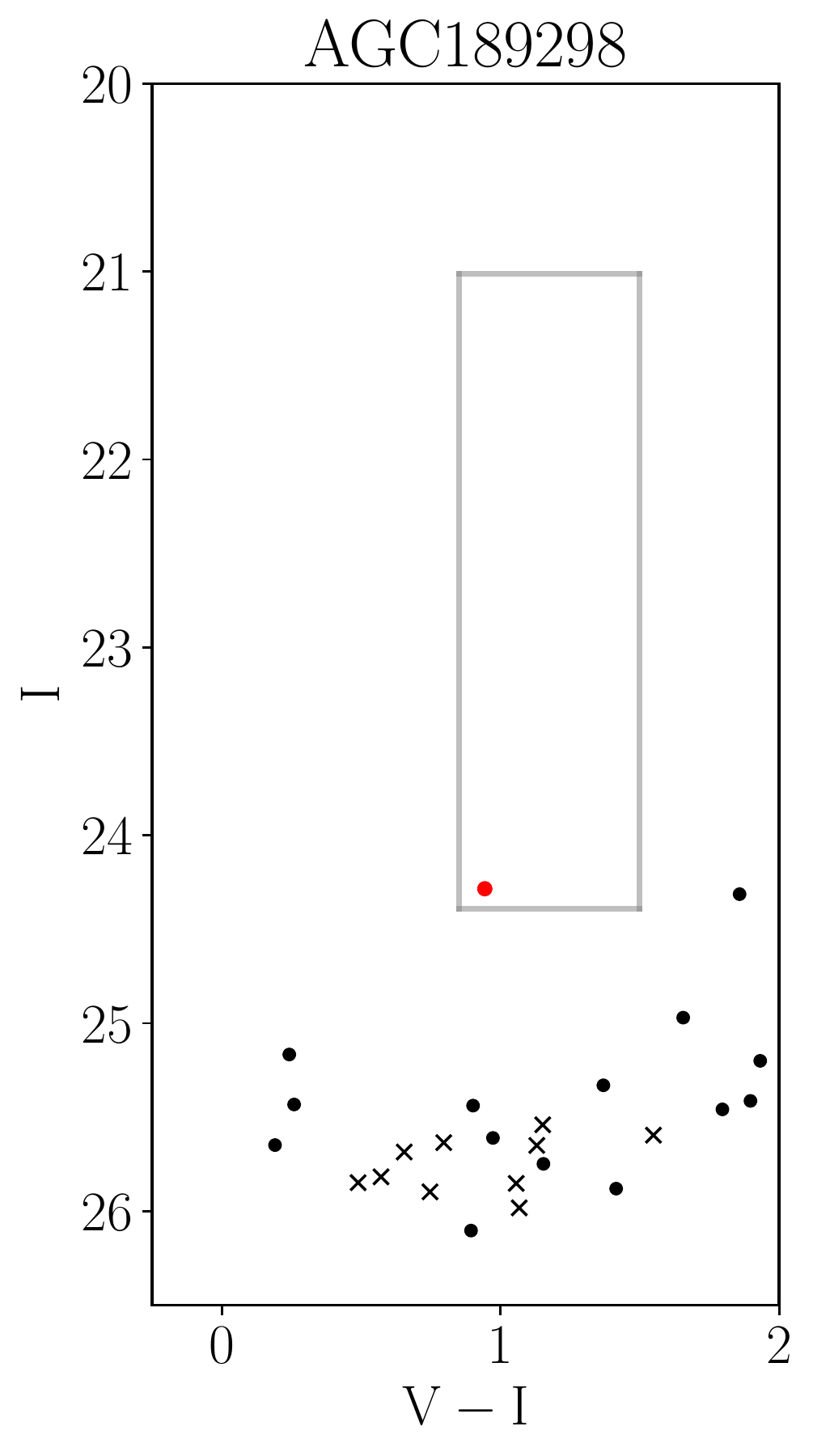}
    \caption{The CMDs of all point-like sources meeting the selection criteria (\S\ref{sec:results}), with $\mathrm{SNR}>5$, and falling within $2r_\mathrm{eff}$ of the center of each target galaxy. Points (red or black) indicate sources that meet the concentration index criterion and black crosses are those that fail it. The grey rectangular outline shows the color-magnitude parameter space used to select GCCs, and red points are those sources that meet all the criteria. }
    \label{fig:CMDs}
\end{figure*}

The CMDs of point-like objects within twice the effective radius of each target galaxy are shown in Figure \ref{fig:CMDs}. 

In most cases there are very few sources in the expected magnitude range of GCs, and neither the color nor concentration index criteria play a significant role in selecting GC candidates. Three exceptions appear to be AGCs 181474, 312297, and 749387. For AGC~181474 there is a  cluster of sources ($0<\mathrm{V-I}<0.75$ and $23 < \mathrm{I} < 25$) in the CMD that are just blueward of the selection box and fail the concentration index criteria. However, these are almost exclusively artifacts in the diffraction spikes of the two bright stars (Figure \ref{fig:HST_images}) near the target galaxy. The same is true for AGC~749387, except the location of the artifacts in CMD is different ($1<\mathrm{V-I}<2$ and $21.5 < \mathrm{I} < 22.5$). In the case of AGC~312297 there are a number of sources in the magnitude range expected for GCs, but just blueward of the selection box. These are likely young star clusters and \hii \ regions, not old GCs.

\begin{figure*}
    \centering
    \includegraphics[width=0.24\textwidth]{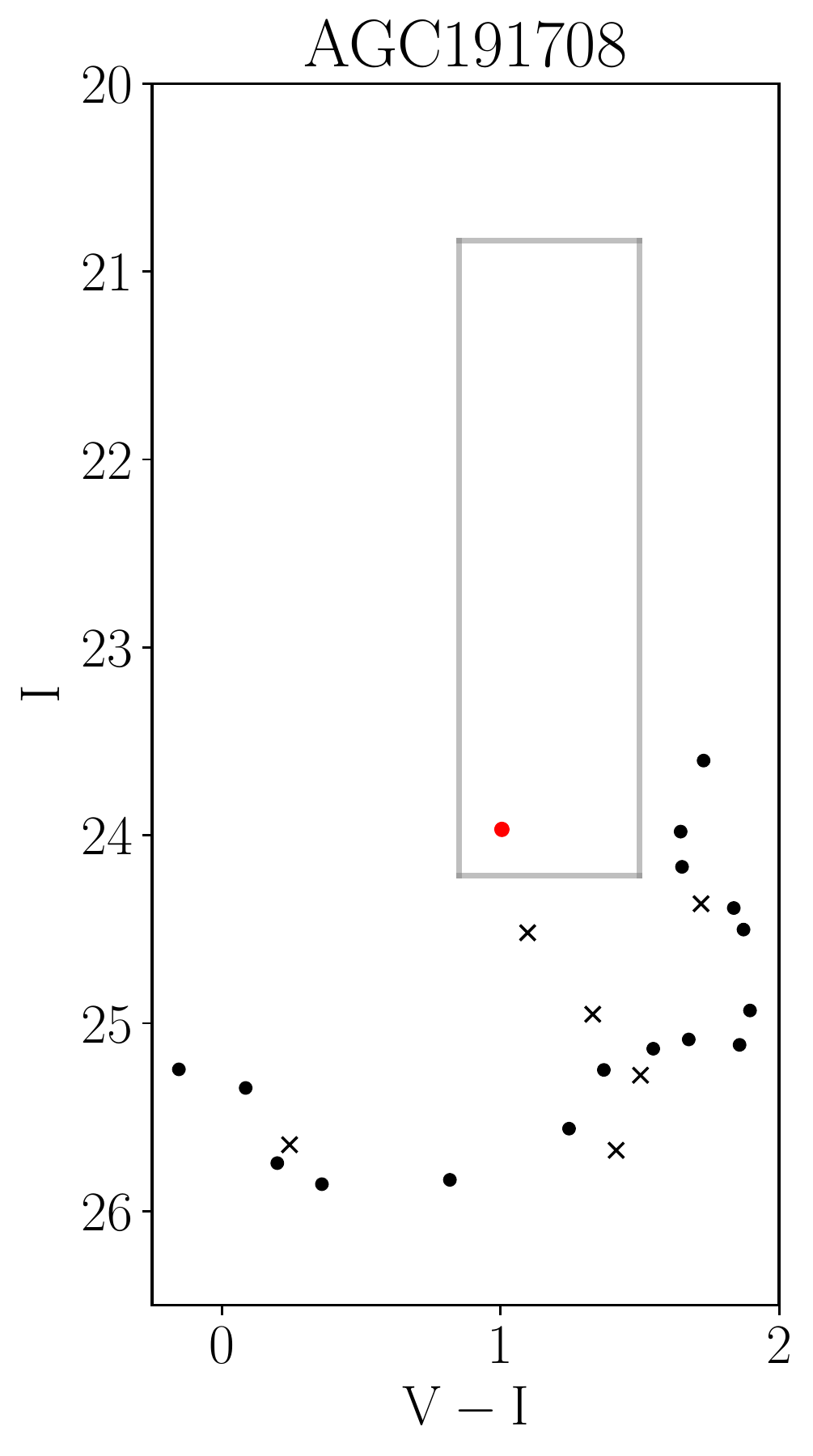}
    \includegraphics[width=0.24\textwidth]{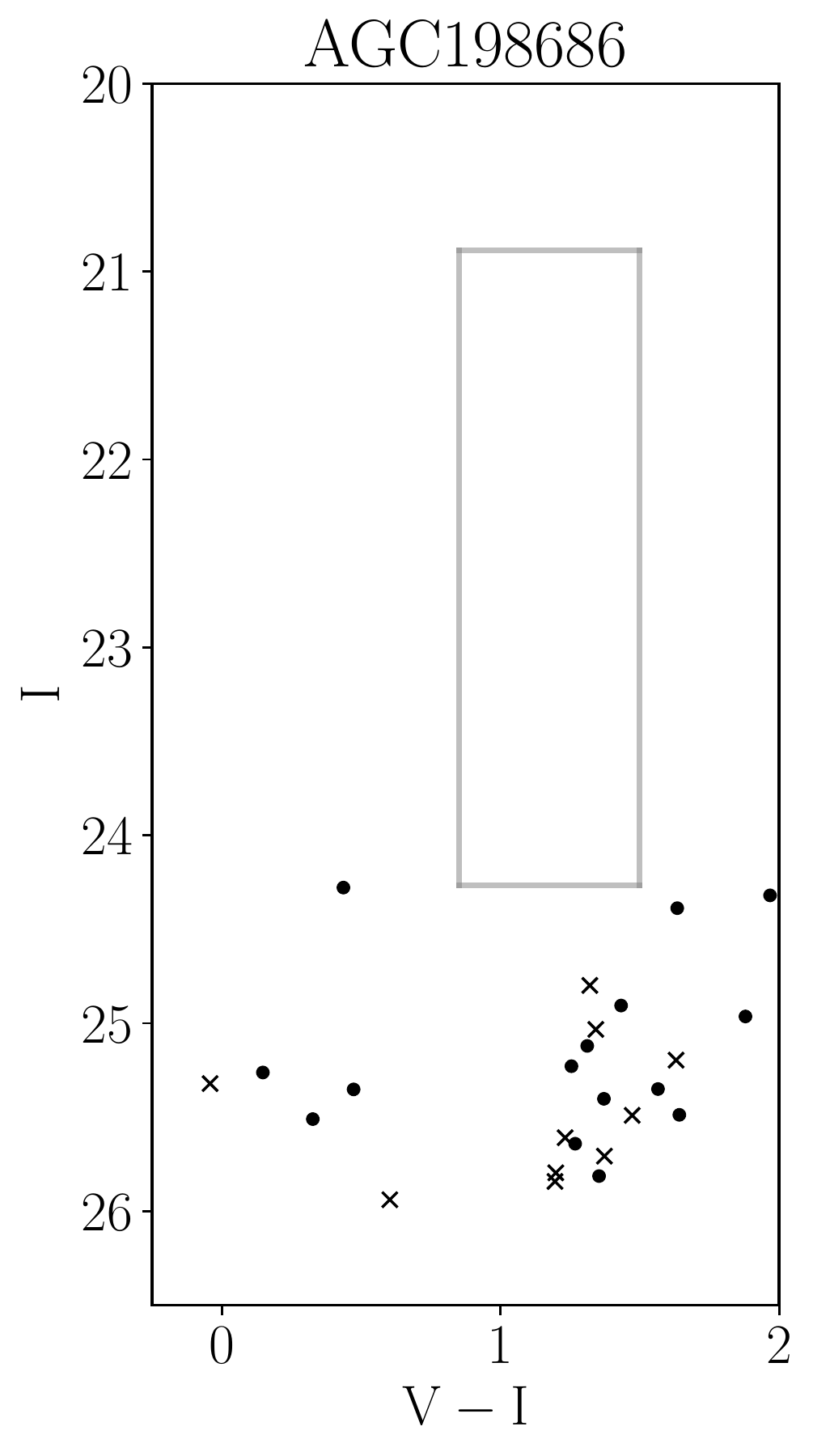}
    \includegraphics[width=0.24\textwidth]{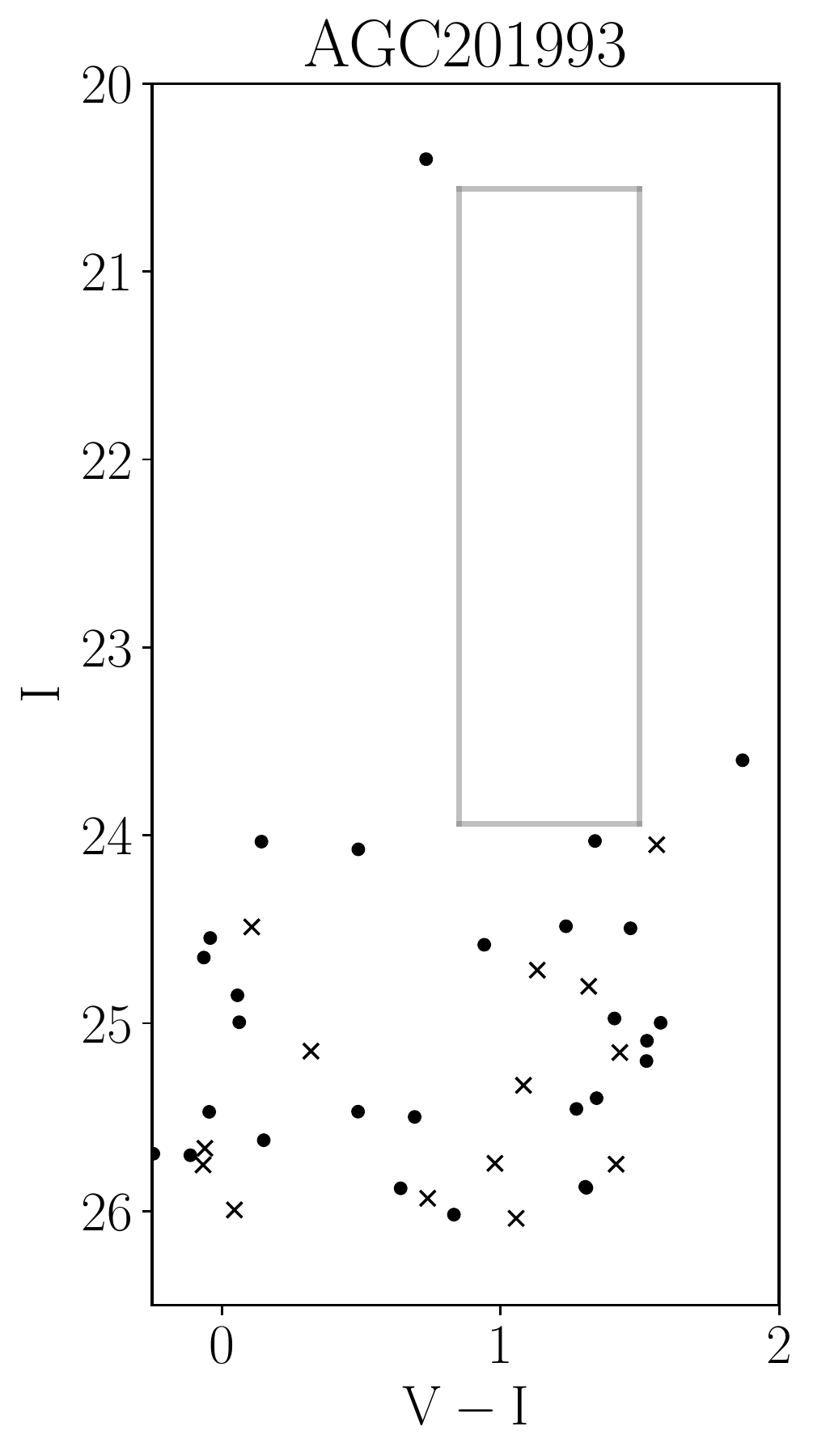}\\
    \centering
    \includegraphics[width=0.24\textwidth]{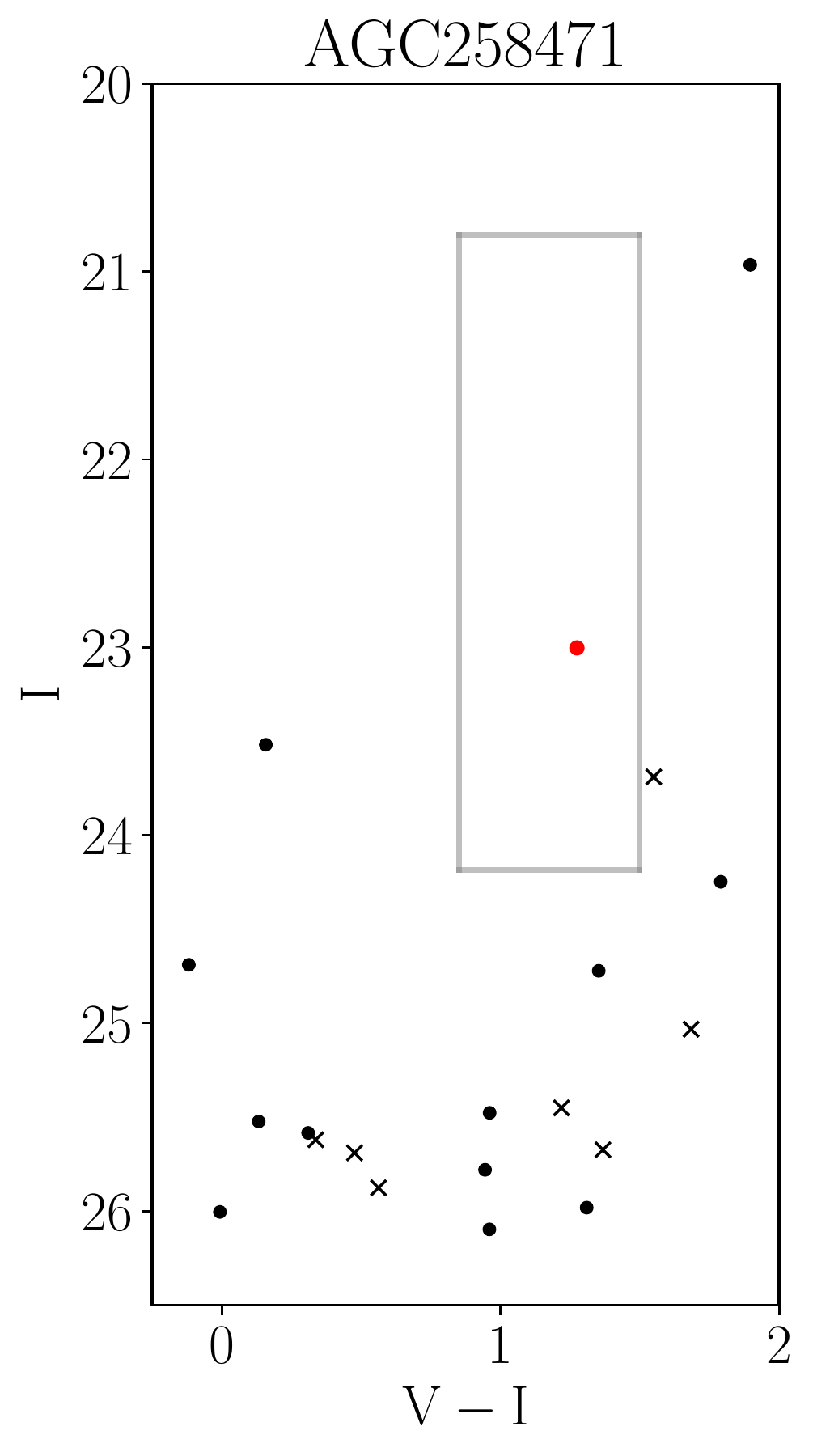}
    \includegraphics[width=0.24\textwidth]{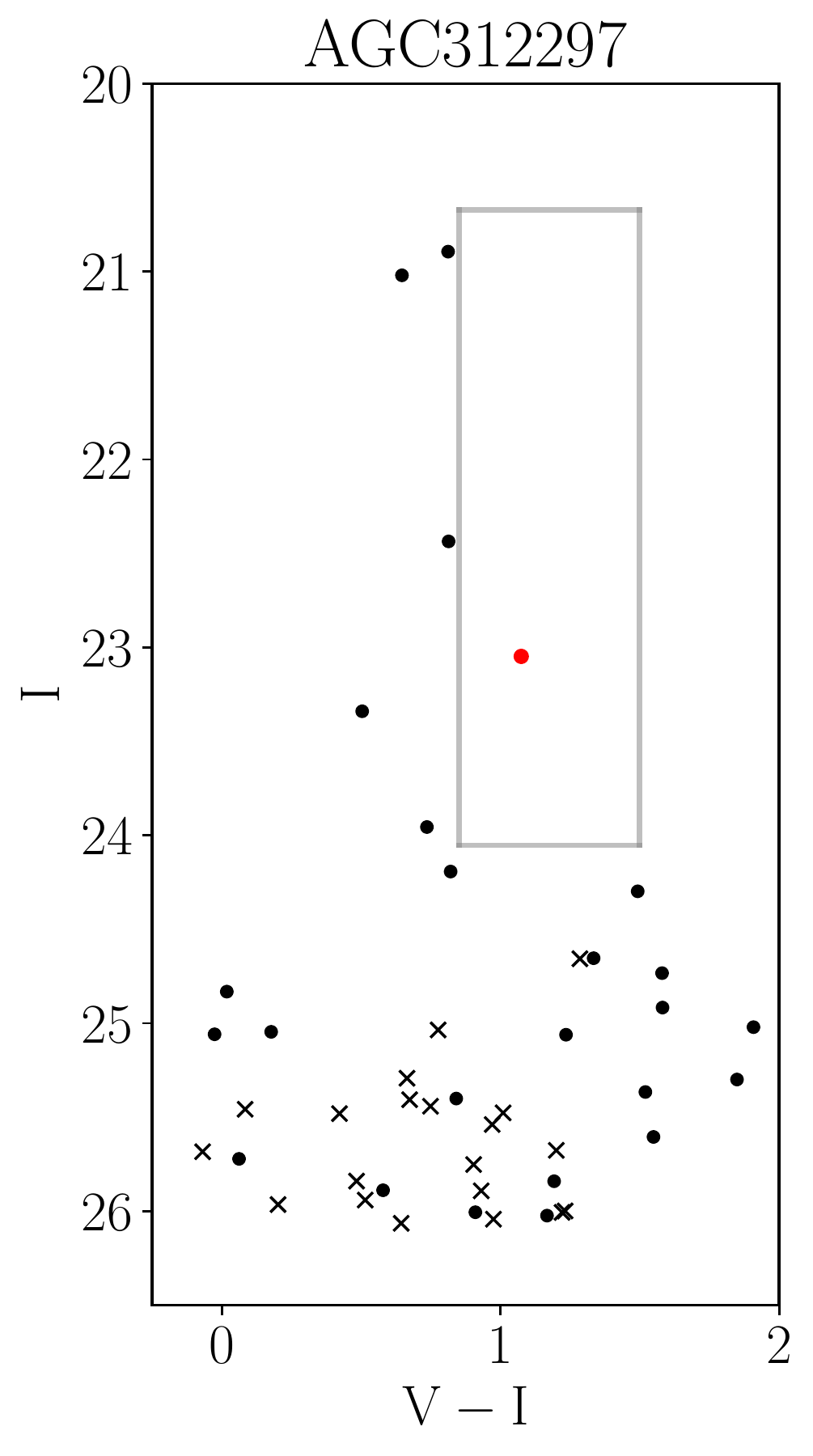}
    \includegraphics[width=0.24\textwidth]{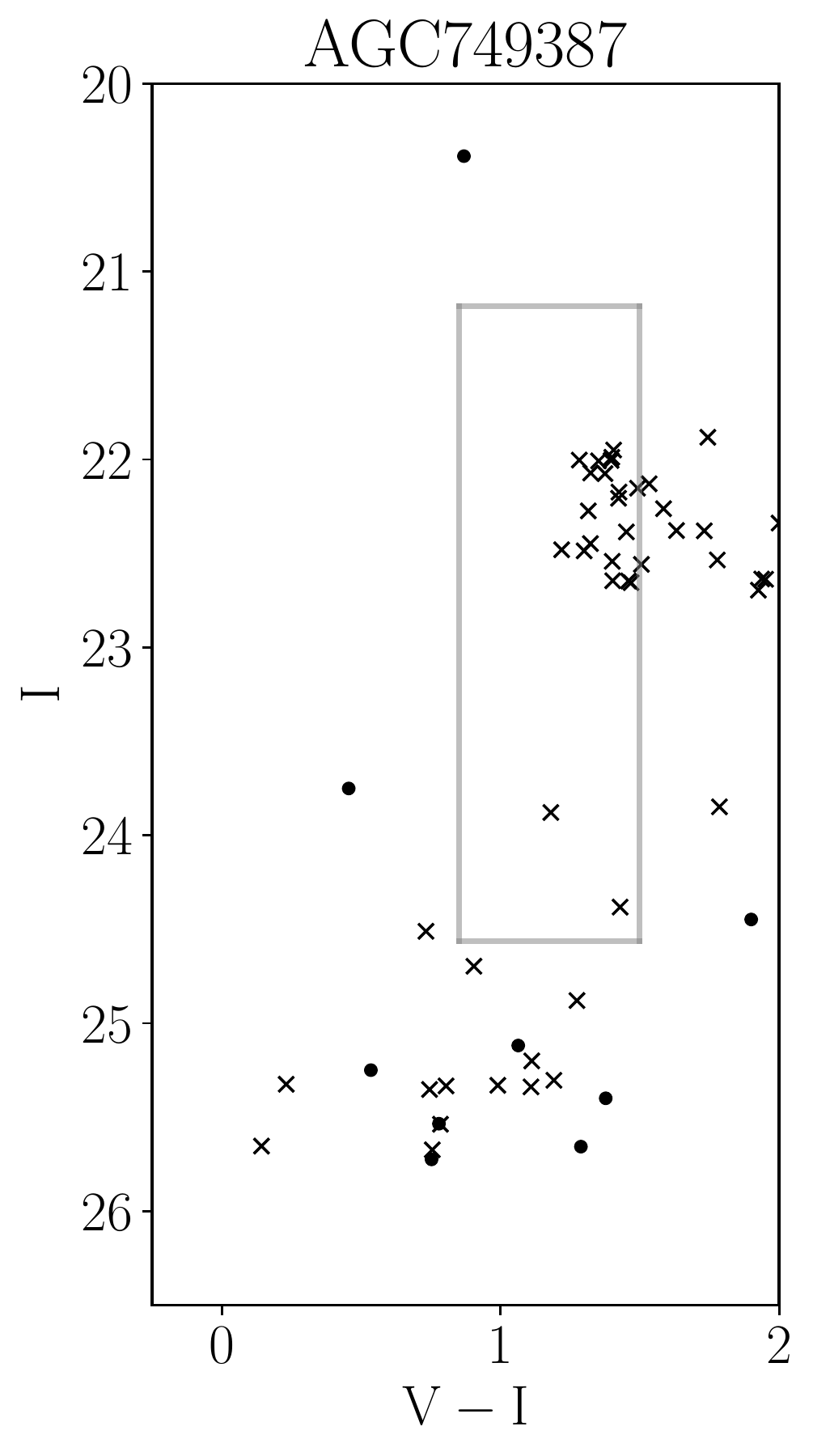}\\
    {\bf Figure \ref{fig:CMDs} continued.}
\end{figure*}

\section{Stellar mass estimates}
\label{sec:stellar_masses}

Given the peculiar nature of our targets we decided to estimate stellar masses using three different relations, all relying on the $g-r$ color and $r$-band absolute magnitude as proxies for the stellar mass. Of the three we found that \citet{Zibetti+2009} consistently gave the lowest mass estimates, \citet{Into+2013}, recalibrated for LSB galaxies by \citet{Du+2020}, gave intermediate mass estimates, and \citet{Herrmann+2016} consistently gave the highest estimates. We therefore elected to use the \citet{Du+2020} recalibration of the \citet{Into+2013} relation as our default choice for stellar mass estimates and used the standard deviation (in dex) between the three methods as the uncertainty estimate.

\bibliography{refs}{}
\bibliographystyle{aasjournal}



\end{document}